\documentclass[11pt,a4paper]{article}

\usepackage[utf8]{inputenc}
\usepackage[T1]{fontenc}
\usepackage{lmodern,amsmath,amssymb,graphicx,authblk,booktabs,subcaption}
\usepackage{float,placeins}
\DeclareCaptionLabelFormat{continuedfigure}{#1~#2 (continued)}
\usepackage[margin=1in]{geometry}
\usepackage{microtype}
\usepackage{hyperref}
\hypersetup{colorlinks=true,linkcolor=blue,citecolor=blue,urlcolor=blue}

\usepackage[noadjust]{cite}

\graphicspath{{figures/}}

\title{\vspace{-0.5em} Phenomenological investigation of order statistics for multi-jet events}

\author[1]{S. Cerci}
\author[2,3]{G.~Chachamis}
\author[4,5]{A.~Sabio Vera}
\author[6]{Z. S. Demiroglu}
\author[1]{D. Sunar Cerci}
\author[7,8]{D.~Vaccaro}

\affil[1]{Department of Physics, Faculty of Arts and Science, Yildiz Technical University, 34220 Esenler, Istanbul, Türkiye}
\affil[2]{Science Department, Universidad Pública de Navarra, Campus de Arrosadia, Pamplona 31006, Spain}
\affil[3]{Institute for Advanced Materials and Mathematics (INAMAT2), Universidad Pública de Navarra, Campus de Arrosadia, Pamplona 31006, Spain}
\affil[4]{Instituto de Física Teórica UAM/CSIC, Nicolás Cabrera 15, E-28049 Madrid, Spain}
\affil[5]{Theoretical Physics Department, Universidad Autónoma de Madrid, E-28049 Madrid, Spain}
\affil[6]{Center for Frontiers in Nuclear Science, Department of Physics and Astronomy, Stony Brook University, Stony Brook, New York 11794-3800, USA}
\affil[7]{Laboratório de Instrumentação e Física Experimental de Partículas (LIP), Av.\ Prof.\ Gama Pinto, 2, P-1649-003 Lisboa, Portugal}
\affil[8]{Departamento de Física, Instituto Superior Técnico, Universidade de Lisboa, Av. Rovisco Pais 1, 1049-001 Lisbon, Portugal}

\date{}

\begin{document}
\maketitle

\begin{abstract}
We investigate rank-ordered jet rapidity distributions as probes of multijet
radiation in proton--proton collisions at $\sqrt{s}=13$~TeV. These observables
retain information about the organization of jets within an event that is
not resolved by the inclusive single-jet rapidity spectrum. We compare
normalized predictions from the BFKL-based generator \textsc{BFKLex} and
the collinear parton-shower generator \textsc{PYTHIA~8.306}, considering both
fixed three-jet configurations and rank observables defined across
multiplicities. We assess PDF and tune variations for \textsc{PYTHIA~8.306},
and PDF and factorization- and renormalization-scale variations for
\textsc{BFKLex}. Total-variation and Wasserstein distances quantify the
separation between the baseline shapes. Differences remain visible relative
to the evaluated uncertainty bands in several well-populated regions.
A minimum rapidity separation between the extremal jets modifies the
discriminating power of the observables in a rank-dependent way.
These results support rank-ordered rapidity spectra as
experimentally accessible observables complementary to conventional
Mueller--Navelet measurements for investigating high-energy QCD radiation.
\end{abstract}

\section{Introduction}

Multijet production at the Large Hadron Collider (LHC) provides a testing ground for the description of radiation in Quantum Chromodynamics (QCD), particularly in the high-energy limit. In this regime, where the center-of-mass energy of the hard scattering is much larger than any other hard scale, the contribution from multigluon emissions compensates the smallness of the strong coupling via large logarithms of energy. These logarithms can be resummed within the Balitsky--Fadin--Kuraev--Lipatov (BFKL) framework~\cite{ Lipatov:1985uk, Balitsky:1978ic, Kuraev:1977fs, Kuraev:1976ge, Lipatov:1976zz, Fadin:1975cb, Fadin:1998py, Ciafaloni:1998gs}.

A variety of processes have been explored to test QCD dynamics in this regime. A prominent example is the production of jets in a Mueller--Navelet (MN) topology~\cite{Mueller:1986ey}, i.e., di-jet events where two high-energy jets are produced with a large separation in rapidity and similar transverse momenta, and the additional jet activity in the event is treated inclusively. A classic probe of this topology is the azimuthal correlation between the forward and backward jets. CMS measured the azimuthal-angle difference, its cosine moments, and ratios of these moments at $\sqrt{s}=7$~TeV, for jets with $p_T>35$~GeV and rapidity separations up to $\Delta y=9.4$~\cite{CMS:2016qng}. A full next-to-leading-logarithmic (NLL) BFKL calculation with Brodsky--Lepage--Mackenzie (BLM) renormalization-scale setting describes these observables within experimental and theoretical uncertainties for $\Delta y>4$~\cite{Ducloue:2013BLM}. However, satisfactory overall agreement is also obtained with HERWIG++, while PYTHIA~6 and PYTHIA~8 describe several of the measured moments. These generators supplement Dokshitzer--Gribov--Lipatov--Altarelli--Parisi (DGLAP) evolution~\cite{Gribov:1972ri, Gribov:1972rt, Altarelli:1977zs, Dokshitzer:1977sg} with parton showers and colour-coherence effects, whose implementation affects the predicted azimuthal decorrelation. Agreement with these measurements therefore does not uniquely identify high-energy resummation effects. This motivates complementary observables that resolve the radiation between the tagging jets more directly, rather than only its net effect on their azimuthal correlation.

In a previous study~\cite{Baldenegro:2024ndr}, some of the present authors and their collaborators explored a more exclusive approach to the characterization of Mueller--Navelet events by introducing a set of observables sensitive to the multi-jet radiation pattern. Comparing predictions from fixed-order QCD supplemented with DGLAP-based parton showers (POWHEG+PYTHIA~8)~\cite{POWHEGRef2,DijetRef, POWHEGRef3,Bierlich:2022pfr} and from the leading logarithmic (LL) BFKL Monte Carlo generator BFKLex~\cite{Chachamis:2015ico, deLeon:2021ecb, Chachamis:2022jis, deLeon:2020myv, Chachamis:2015zzp, Caporale:2013bva, Chachamis:2012qw, Chachamis:2012fk, Chachamis:2011nz, Chachamis:2011rw}, they found that, under realistic LHC conditions, the proposed event-shape observables exhibited only limited sensitivity to the underlying dynamics. This suggested that even relatively exclusive observables may still average over part of the information carried by the radiation pattern, motivating the search for new observables capable of retaining a larger fraction of the event structure.

In Ref.~\cite{Chachamis:2026ostats}, some of us introduced rank-ordered jet rapidity distributions as probes of multijet radiation at $\sqrt{s}=13$~TeV. Comparisons of BFKLex with PYTHIA~8 and HERWIG~7 revealed shape differences that can be hidden in inclusive spectra: BFKLex populates the rapidity interval more evenly, whereas the collinear showers exhibit stronger edge enhancement and central concentration. These differences, quantified through total-variation and Wasserstein distances, persist under harder event selections.

Building on these findings, we investigate whether the separation between \textsc{BFKLex} and \textsc{PYTHIA~8.306} predictions remains robust once the uncertainties considered in this study are taken into account. We examine normalized rapidity distributions both at fixed jet multiplicity and for rank observables defined across multiplicities. For \textsc{PYTHIA~8.306}, we assess the effects of PDF and tune variations, combining the corresponding uncertainty envelopes with the nominal Monte Carlo statistical uncertainty. For \textsc{BFKLex}, we study variations of the factorization and renormalization scales, $\mu_F$ and $\mu_R$, as well as PDF variations.

The remainder of this paper is organized as follows. In Sec.~2 we introduce the order-statistics framework and establish the connection between unordered rapidity distributions and the corresponding rank-ordered observables, illustrating the formalism with simple reference examples before extending it to realistic collider events. Section~3 reviews the high-energy QCD framework and the BFKL formalism, introducing the BFKLex Monte Carlo generator used throughout this work. The phenomenological setup, including the event selection, the observables under study, and the treatment of theoretical and Monte Carlo uncertainties is described in Sec.~4. The main phenomenological results are discussed in Sec.~5. Finally, Sec.~6 summarizes our conclusions and outlines possible future developments.

\section{Order statistics for jet rapidity observables}
\label{sec:orderstats}

In this section we briefly review the statistical
concept of order statistics, applied to jet observables. Assuming independent
and identically distributed (i.i.d.) rapidities, we derive the mathematical
mapping between a parent rapidity distribution and the distributions of the
ordered jet rapidities. The i.i.d. assumption is an idealization of real
multijet events, whose jet rapidities are correlated. Nevertheless, it
provides a useful theoretical starting point for isolating the effects of
ordering and an analytic reference for the rank observables studied below.

\subsection{Order Statistics}
Consider first a fixed-multiplicity event sample. Each event contains exactly
$N$ accepted jets with unordered rapidities
\begin{equation}
  Y_1,Y_2,\ldots,Y_N .
\end{equation}
As an analytic reference, assume that these variables are independent and
identically distributed with continuous parent density $f(y)$ and cumulative
distribution
\begin{equation}
  F(y)=\Pr(Y_i\leq y)=\int_{-\infty}^{y} f(u)\,du .
\end{equation}
Continuity implies that ties occur with zero probability, so every event
admits a unique ordering
\begin{equation}
  Y_{(1)}<Y_{(2)}<\cdots<Y_{(N)} .
\end{equation}
The variable $Y_{(k)}$ occupying position $k$ is the $k$th
\emph{rank-ordered} jet. The distribution of $Y_{(k)}$ follows directly from a counting argument.
For a fixed rapidity $y$, define
\begin{equation}
  K(y)=\sum_{i=1}^{N}\mathbf{1}_{\{Y_i\leq y\}},
\end{equation}
the number of observations lying at or below $y$. Each indicator has success
probability $F(y)$, therefore
\begin{equation}
  K(y)\sim{\rm Binomial}\!\left(N,F(y)\right).
\end{equation}
Now $Y_{(k)}\leq y$ if and only if at least $k$ of the $N$ observations
are no greater than $y$. Hence the cumulative distribution of the $k$th
order statistic is given by a binomial upper-tail probability
\begin{equation}
  G_{k}(y)
  \equiv \Pr\!\left(Y_{(k)}\leq y\right)
  =\sum_{j=k}^{N}
  \binom{N}{j}[F(y)]^{j}[1-F(y)]^{N-j}.
  \label{eq:order-cdf}
\end{equation}
Differentiating it with respect to
$y$ and using $F'(y)=f(y)$, the sum telescopes and all interior terms cancel,
leaving only the boundary contribution. This yields
\begin{equation}
  f_{(k)}(y)=
  \frac{N!}{(k-1)!(N-k)!}\,
  [F(y)]^{k-1}[1-F(y)]^{N-k}f(y),
  \qquad k=1,\ldots,N .
  \label{eq:orderstats}
\end{equation}
The factors in this expression have an immediate interpretation: one
observation lies in an infinitesimal interval around $y$, exactly
$k-1$ lie below it, and the remaining $N-k$ lie above it. The
factorial coefficient counts the possible assignments of the $N$
observations to these three groups.

The probability-integral transform makes the result distribution-free.
Defining $X_i=F(Y_i)$ produces independent uniform variables on $[0,1]$, and
monotonicity gives $X_{(k)}=F(Y_{(k)})$. Equation~\eqref{eq:orderstats}
therefore reduces to
\begin{equation}
  f_{X_{(k)}}(x)=
  \frac{N!}{(k-1)!(N-k)!}\,
  x^{k-1}(1-x)^{N-k},
  \qquad 0\leq x\leq 1,
\end{equation}
so that
\begin{equation}
  X_{(k)}\sim{\rm Beta}(k,N-k+1),\qquad
  \mathbb{E}[X_{(k)}]=\frac{k}{N+1},\qquad
  {\rm Var}(X_{(k)})=
  \frac{k(N-k+1)}{(N+1)^2(N+2)} .
\end{equation}
For a uniform rapidity parent on $[-Y/2,Y/2]$, one simply has
$x=(y+Y/2)/Y$. Thus, even when the unordered rapidities are completely flat,
the edge ranks accumulate near the interval boundaries while the central
ranks concentrate progressively towards its middle.

\vspace{0.5em}

\begin{figure}[!t]
  \centering
  \begin{subfigure}[t]{0.49\textwidth}
    \centering
    \includegraphics[width=\linewidth,trim=0 0 80 14,clip]{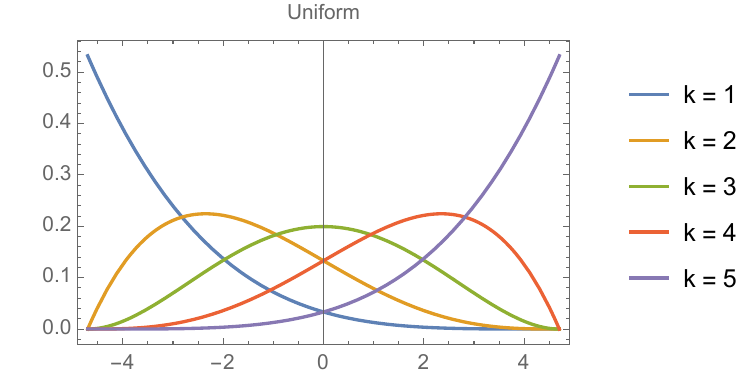}
    \caption{Uniform parent.}
    \label{fig:orders-uniform}
  \end{subfigure}\hfill
  \begin{subfigure}[t]{0.49\textwidth}
    \centering
    \includegraphics[width=\linewidth,trim=0 0 80 14,clip]{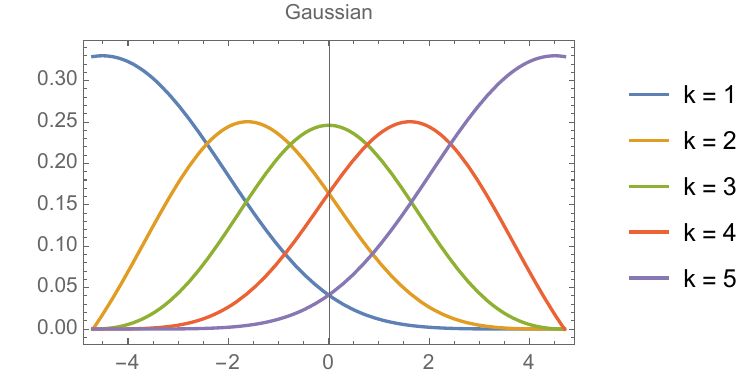}
    \caption{Gaussian parent $\mathcal{N}(0,16)$.}
    \label{fig:orders-gaussian}
  \end{subfigure}

  \smallskip
  \begin{subfigure}[t]{0.49\textwidth}
    \centering
    \includegraphics[width=\linewidth,trim=0 0 80 14,clip]{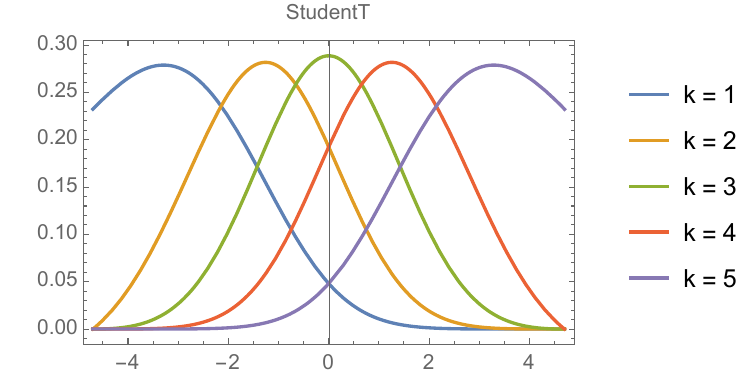}
    \caption{Student-$t$: $\mu=0$, $\sigma=3$, $\nu=5$.}
    \label{fig:orders-studentt}
  \end{subfigure}
  \hspace{0.02\textwidth}%
  \begin{minipage}[b]{0.11\textwidth}
    \centering
    \includegraphics[width=\linewidth,trim=300 36 0 38,clip]{studentT-rank-5.pdf}
    \vspace{1.5em}
  \end{minipage}
  \caption{Order-statistic densities $f_{(k)}(y)$ for $N=5$ independent jets
  drawn from three parent densities restricted to $[-4.7,4.7]$.
  The Gaussian variance is $16$ before truncation; $\sigma$ denotes the
  Student-$t$ scale parameter. Ordering a uniform parent produces distinct
  rank shapes; increasing the central concentration moves the outer ranks
  inwards and narrows the central rank distribution.}
  \label{fig:orders-parents}
\end{figure}

To illustrate how the parent density affects the rank distributions, we consider exactly $N=5$ jets drawn independently from uniform, Gaussian, and Student-$t$ densities restricted to $\mathcal{W}=[-4.7,4.7]$. For the uniform parent distribution, Fig.~\ref{fig:orders-uniform}, the probability density is given by
\begin{equation}
  f(y)=\frac{1}{y_{\max}-y_{\min}},
  \qquad y\in\mathcal{W}.
\end{equation}
The extreme ranks accumulate near opposite
boundaries, while the inner ranks peak progressively closer to the centre.
All of this structure is generated by the ordering itself, since every
unordered rapidity bin is equally probable.

For a Gaussian produced density $\mathcal{N}(0,16)$ with mean zero and variance $16$, Fig.~\ref{fig:orders-gaussian}, truncated to the detector
window, the accepted parent is moderately enhanced at central rapidities.
The ordered densities are consequently pulled towards $y=0$: the central
ranks become more localized, while the extreme ranks retain less weight near
the boundaries than in the uniform case.

A Student-$t$ density with location $\mu=0$, scale $\sigma=3$, and $\nu=5$ degrees of freedom, Fig.~\ref{fig:orders-studentt}, again truncated to
$\mathcal{W}$, is more strongly concentrated around the centre for the
parameters used here. The inner ranks become narrower and the extreme ranks
are further suppressed near the acceptance boundaries.

The discussion above shows that, for a fixed number of i.i.d. rapidities
drawn from any continuous parent distribution, the transformation
$X_{(k)}=F(Y_{(k)})$ yields a
$\mathrm{Beta}(k,N-k+1)$ distribution. In the original rapidity
variable, however, the rank shapes depend on the parent density, as
illustrated in Fig.~\ref{fig:orders-parents}. The uniform parent gives
rescaled Beta densities across the acceptance interval. For the truncated
Gaussian parent considered here, the outer ranks move towards the centre
and the central ranks become more localized. The chosen truncated
Student-$t$ parent is more centrally concentrated and further enhances
these features. Thus, the universal Beta form in the cumulative variable
coexists with distinct rank shapes in rapidity space.

\subsection{Order Statistics for jet observables}
\label{sec:collider-interpretation}
The examples above show that we can use order statistics as a fingerprint for the parent density distribution in jet events. This new class of observables could provide more exclusive information than the inclusive ones. To see what ordering recovers, and what information is missing in inclusive observables, consider the normalized single-inclusive rapidity spectrum,
\begin{equation}
  \rho_1(y)=\frac{1}{N_{\mathrm{jets}}^{\mathrm{total}}}\,
  \frac{dN_{\mathrm{jets}}}{dy} .
\end{equation}
Here $N_{\mathrm{jets}}^{\mathrm{total}}$ is the total number of jets passing the fiducial
selection in the selected event sample. The spectrum sums over accepted
multiplicities and event configurations and is normalized to unity,
\begin{equation}
  \int_{\mathcal{W}}\rho_1(y)\,dy=1.
\end{equation}
It is therefore the probability density for the rapidity of a jet drawn
from the accepted jet sample. It records where jets occur, but not which
other jets occur in the same event. Overall multiplicative rate factors
cancel in this normalization, whereas PDFs, underlying-event activity,
and hadronization can still affect the shape.

A two-jet example makes the missing information explicit. In Model~A, the
rapidities are independent and uniform in a symmetric window. In Model~B,
they are constrained to the pair $(-U,+U)$, with $U$ uniform in the forward
half of the window. Both models have the same uniform $\rho_1(y)$, yet their
event structures are different. Model~A follows the independent
order-statistics construction, whereas Model~B satisfies
$Y_{(1)}=-U$ and $Y_{(2)}=+U$ event by event. The distributions of the two
ranks and of their separation therefore distinguish the models immediately.
Thus, although rank distributions remain ordinary one-dimensional
histograms, they retain information about how the complete set of rapidities
is organized within each event.

This sensitivity provides the physical bridge to high-energy scattering.
In a multiperipheral description, produced objects are strongly ordered in
rapidity and only weakly ordered in transverse momentum, leading in the
idealized limit to an approximately uniform population of the available
rapidity interval. Once two boundary jets define an interval of length $Y$,
the ordered positions of the jets describe directly how that
interval is filled, and Eq.~\eqref{eq:orderstats} converts the qualitative
picture into a quantitative reference template. BFKL ladders inherit the
strong rapidity ordering of multi-Regge kinematics while allowing diffusion
in transverse momentum, and can therefore populate the inter-tag region
broadly. Collinear parton showers provide a different description of the radiation
pattern. Differences between the two approaches can appear as rank-dependent
shape deformations, motivating the use of these observables to discriminate
between predictions based on high-energy resummation and collinear parton
showers.

However, to connect the statistical templates to measured jets, we need to face 
the limitations dictated by the experimental setup, and the parent density
must be defined after the fiducial selection. Only jets inside the detector
rapidity window $\mathcal{W}=[y_{\min},y_{\max}]$ are observed, and in this
analysis $y_{\min}=-4.7$ and $y_{\max}=+4.7$. If the produced one-particle
density is $f_{\rm prod}(y)$ on the real line, conditioning on the acceptance
gives
\begin{equation}
  f_{\rm obs}(y)=
  \frac{f_{\rm prod}(y)}
       {\displaystyle\int_{y_{\min}}^{y_{\max}}f_{\rm prod}(u)\,du},
  \qquad y\in\mathcal{W}.
  \label{eq:truncated}
\end{equation}
The transverse-momentum threshold can add a second conditioning effect. If
rapidity and transverse momentum are independent for each jet, imposing
$p_T>p_T^{\rm cut}$ leaves $f_{\rm obs}(y)$ unchanged. In general, however,
the effective accepted density is
\begin{equation}
  f_{\rm obs}(y)\propto
  f_{\rm prod}(y)\,
  \Pr\!\left(p_T>p_T^{\rm cut}\mid Y=y\right),
  \qquad y\in\mathcal{W},
  \label{eq:reweight}
\end{equation}
so a rapidity-dependent selection efficiency is inherited by every ordered
spectrum.

Equation~\eqref{eq:orderstats} should therefore be read as an interpretive
benchmark, not as an assumption that collider jets are independent.
In realistic multijet events, QCD radiation generates correlations through
color coherence, while energy-momentum conservation and recoil constrain
the emission kinematics. These correlations are part of the physical
information that rank-ordered observables can probe beyond the i.i.d.
reference. Their manifestation in jet distributions also depends on the
jet definition, including the radius parameter and recombination scheme,
and on the fiducial selection. In particular, transverse-momentum cuts
can reshape the accepted rapidity density as in Eq.~\eqref{eq:reweight},
while the rapidity window determines which jets enter the ordering.
Comparisons between theoretical predictions should therefore use a common
jet definition and selection. For comparisons with data, detector effects
must additionally be accounted for by applying the detector response to
the predictions.

\FloatBarrier

\section{BFKL dynamics and \textsc{BFKLex} generator}
\label{sec:bfkl-bfklex}

The BFKL formalism resums logarithmically enhanced contributions in the
high-energy limit of QCD, where the partonic center-of-mass energy
$\sqrt{\hat{s}}$ is much larger than the transverse hard scales.
The relevant expansion parameter is $\alpha_s\ln(\hat{s}/s_0)$, where
$s_0$ is a reference energy scale with dimensions of energy squared,
chosen of the order of the product of the transverse hard scales.
For two tagged objects A and B separated by $\Delta Y=y_A-y_B$,
$\ln(\hat{s}/s_0)\simeq\Delta Y$ in the high-energy limit with an
appropriate choice of $s_0$. The contributions
$\alpha_s^n(\Delta Y)^n$ can therefore become numerically important
and require resummation.

Following the presentation in Ref.~\cite{Baldenegro:2024ndr}, we summarize
the logarithmic counting and the iterative formulation underlying the
BFKLex calculation.
From a logarithmic power-counting perspective, longitudinal and transverse degrees of freedom decouple at LL accuracy, yielding the schematic series
\begin{equation}
  \sigma_{\text{LL}} \;=\; \sum_{n=0}^{\infty} C^{\text{LL}}_{n}\,\alpha_s^n
  \int_{y_B}^{y_A}\! dy_1 \int_{y_B}^{y_1}\! dy_2 \cdots \int_{y_B}^{y_{n-1}}\! dy_n
  \;=\; \sum_{n=0}^{\infty} \frac{C^{\text{LL}}_{n}}{n!}\,\alpha_s^n\,(y_A-y_B)^n,
  \label{eq:ll-series}
\end{equation}
where $y_i$ denote the rapidities of emitted particles and the coefficients $C^{\text{LL}}_{n}$ are calculable within LL BFKL resummation. At NLL accuracy the resummation becomes sensitive to the running of $\alpha_s$ and to the energy-scale choice that enters the logarithms. A convenient parameterization of these scheme dependences introduces constants $A$ and $B$ in the LL expression, illustrating how scale choices generate terms with one fewer high-energy logarithm at a given power of $\alpha_s$,
\begin{gather}
\sigma^{\text{LL+NLL}}
=
\sum_{n=0}^{\infty}
\frac{C_n^{\text{LL}}}{n!}
(\alpha_s-A\alpha_s^2)^n
(y_A-y_B-B)^n \nonumber \\
= \sigma^{\text{LL}} - \sum_{n=1}^{\infty} \frac{B C_n^{\text{LL}} + (n-1)AC_{n-1}^{\text{LL}}}{(n-1)!}\, \alpha_s^n (y_A - y_B)^{n-1}+ ... \ .
\end{gather}
This schematic expansion illustrates scale-induced NLL contributions; it
is not a substitute for the full NLL corrections to the kernel and the
process-dependent impact factors.

The corresponding factorized representation of the hard cross section reads
\begin{equation}
  \sigma(Q_1,Q_2,\Delta Y)
  \;=\;
  \int d^2\vec{k}_A\, d^2\vec{k}_B\;
  \underbrace{\vphantom{\int}\phi_A(Q_1,\vec{k}_A)\,\phi_B(Q_2,\vec{k}_B)}_{\text{process-dependent impact factors}}
  \;\underbrace{f(\vec{k}_A,\vec{k}_B,\Delta Y)}_{\text{universal BFKL Green's function}},
  \label{eq:fact}
\end{equation}
where the impact factors $\phi_{A,B}$ encode the short-distance process (including external hard scales $Q_{1,2}$) and the BFKL Green's function $f$ is universal. In momentum space, the LL BFKL equation for the $\omega$-Mellin transform of $f$ reads
\begin{equation}
  \omega\, f_{\omega}(\vec{k}_A,\vec{k}_B)
  \;=\; \delta^{(2)}(\vec{k}_A-\vec{k}_B) \;+\;
  \int d^2\vec{k}\; K(\vec{k}_A,\vec{k})\,f_{\omega}(\vec{k},\vec{k}_B),
  \label{eq:bfkl-eq}
\end{equation}
with kernel
\begin{equation}
  K(\vec{k}_a,\vec{k})
  \;=\;
  2\,\omega(\vec{q})\,\delta^{(2)}(\vec{k}_a-\vec{k})
  \;+\;
  \frac{N_c\,\alpha_s}{\pi^2}\,\frac{1}{(\vec{k}_a-\vec{k})^{2}},
  \label{eq:kernel}
\end{equation}
where $\vec q=\vec k_a$ denotes the transverse momentum entering the virtual term. The first term corresponds to virtual corrections via the gluon Regge trajectory $\omega(\vec{q})$, and the second to real emissions.

An iterative (ladder) solution at LL can be written in transverse momentum space as a Sudakov-like exponential of the trajectory multiplied by a sum over emissions with strongly ordered rapidities~\cite{PhysRevLett.78.4531}. It can be expressed as
\begin{multline}
f
=
e^{\omega(\vec{k}_A)\Delta Y}
\Bigg[
\delta^{(2)}(\vec{k}_A-\vec{k}_B)
+
\sum_{n=1}^{\infty}
\prod_{i=1}^{n}
\frac{\alpha_s N_c}{\pi}
\int
\frac{d^2\vec{k}_i}{\pi k_i^2}
\theta(k_i^2-\lambda^2)
\\
\times \int_0^{y_{i-1}}dy_i
\,
e^{(\omega(\vec{k}_A+\sum_{l=1}^i \vec{k}_l)-\omega(\vec{k}_A+\sum_{l=1}^{i-1} \vec{k}_l))y_i}
\,
\delta^{(2)}
\left(
\vec{k}_A
+
\sum_{l=1}^{n}\vec{k}_l
-
\vec{k}_B
\right)
\Bigg],
\end{multline}
where $y_0=\Delta Y$ and the emission rapidities are measured relative to the lower tagging rapidity $y_B$, so that $0<y_n<\cdots<y_1<\Delta Y$. The parameter $\lambda$ is an infrared regulator, and the gluon Regge trajectory is given by
\begin{equation}
\omega(\vec{q})
=
-
\frac{\alpha_s N_c}{\pi}
\log\left(\frac{q^2}{\lambda^2}\right).
\end{equation}
The Monte Carlo event generator \textsc{BFKLex} provides the exclusive
realization needed for this purpose. In the LL setup used in this work,
the calculation uses the iterative solution of the BFKL Green's function
in transverse momentum space, combines it with the appropriate jet vertices
(impact factors), and convolves the resulting partonic cross section with
parton distribution functions~\cite{Baldenegro:2024ndr}. Each generated event is represented by a
list of final-state gluons with four-momenta, which can be clustered into jets
and analyzed with standard algorithms such as anti-$k_t$. The resulting
partonic final states preserve the characteristic rapidity ordering of
high-energy evolution without imposing strong transverse-momentum ordering.

\section{Setup}
We compare predictions from \textsc{BFKLex} and \textsc{PYTHIA~8.306}, using the latter's $p_T$-ordered parton shower. Jets are reconstructed from the final-state partons using the anti-$k_t$ algorithm implemented in \textsc{FastJet}~\cite{Cacciari:2008gp,Cacciari:2011ma} and are then ordered by rapidity as described above. For \textsc{PYTHIA~8.306}, we assess PDF and tune variations. For \textsc{BFKLex}, we study variations of the factorization and renormalization scales, $\mu_F$ and $\mu_R$, as well as PDF variations.

For \textsc{PYTHIA~8.306}, we use a configuration with multiparton interactions (MPI) enabled and hadronization disabled. As already stated in our previous study~\cite{Chachamis:2026ostats}, hadronization has only a small effect on the normalized rank-ordered rapidity distributions.

\subsection{Observables and kinematics}

We study multijet events in which all accepted jets are reconstructed and ranked using the same criteria. The most-backward and most-forward jets are identified by their positions in the rapidity ordering.

\begin{samepage}
\noindent We generate events with the following setup:
\begin{enumerate}
  \item Proton--proton collisions at $\sqrt{s}=13$~TeV.
  \item A parton-level generation cut \texttt{PhaseSpace:pTHatMin = 20.0}~GeV in \textsc{PYTHIA~8.306} and $k_a, k_b \geq 20$~GeV for the two partonic endpoints of the BFKL ladder in BFKLex.
  \item Rapidity acceptance $-4.7<y<4.7$ for all jets.
  \item Jets reconstructed with the anti-$k_t$ algorithm and radius parameter $R=0.4$.
  \item An analysis-level jet threshold $p_T>30$~GeV.
  \item All accepted jets are ordered by rapidity.
\end{enumerate}
\end{samepage}

To avoid a completely symmetric boundary-jet selection while retaining both
orientations of the event, we impose the union of the two asymmetric cuts
\begin{equation}
 \left(p_{T,\mathrm{MB}}>30~\mathrm{GeV}
       \;\land\;p_{T,\mathrm{MF}}>60~\mathrm{GeV}\right)
 \;\lor\;
 \left(p_{T,\mathrm{MF}}>30~\mathrm{GeV}
       \;\land\;p_{T,\mathrm{MB}}>60~\mathrm{GeV}\right),
 \label{eq:asymmetric-boundary-cuts}
\end{equation}
where MB and MF denote the most-backward and most-forward boundary jets,
respectively. 

\medskip
\noindent
Moreover, results are produced both without an additional rapidity-gap
requirement, $\Delta Y_{\mathrm{MF,MB}}\equiv
y_{\mathrm{MF}}-y_{\mathrm{MB}}\geq 0$, and with
$\Delta Y_{\mathrm{MF,MB}}\geq 3$. The latter selection moves towards a Mueller--Navelet topology with
widely separated boundary jets and allows us to study how the rank
observables depend on the rapidity-gap requirement.

\medskip
\noindent
We analyze two families of histograms:
\begin{enumerate}
  \item \textbf{Fixed multiplicity $N=3$:} the normalized rapidity distributions for three jets ordered by rapidity, from most backward to most forward.
  \item \textbf{Across multiplicities:} the normalized rapidity distributions for the most-forward (MF) and most-backward (MB) jets, and the second-most-forward (SMF) and second-most-backward (SMB) jets, collected from events with $N\geq2$ accepted jets. Each event contributes one entry to each rank distribution; for $N=2$, SMB coincides with MF and SMF with MB.
\end{enumerate}
Both sets characterize the rapidity structure through jet ranks, using the order-statistics framework introduced above, and are straightforward to measure. They complement canonical Mueller--Navelet observables such as azimuthal decorrelations by probing how jets populate the rapidity interval through simple one-dimensional distributions.

\subsection{Integral distances between distributions}
\label{sec:shape-distances}

To quantify the separation between the two normalized histograms, we use
complementary shape distances following Ref.~\cite{Chachamis:2026ostats}.
For a common binning, let $h_i$ and $g_i$ be the normalized
\textsc{PYTHIA~8.306} and \textsc{BFKLex} densities in bin $i$, respectively.
The corresponding bin probabilities are $p_i=h_i\Delta y_i$ and
$q_i=g_i\Delta y_i$, with $\sum_i p_i=\sum_i q_i=1$.
The total-variation distance~\cite{Gibbs:2002metrics},
\begin{equation}
  \mathrm{TV}(p,q)=\frac{1}{2}\sum_{i=1}^{N_{\mathrm{bin}}}|p_i-q_i|,
  \label{eq:total-variation-distance}
\end{equation}
measures the fraction of probability weight that must be redistributed
among bins. It is dimensionless and lies between zero, for identical
histograms, and one, for histograms with disjoint support.

The one-dimensional Wasserstein distance~\cite{Gibbs:2002metrics} also accounts for the separation
in rapidity between bins,
\begin{equation}
  W_1(p,q)=\int dy\,|F_p(y)-F_q(y)|,
  \label{eq:wasserstein-distance}
\end{equation}
where $F_p$ and $F_q$ are the cumulative distributions. Using bin centres
$y_i$, its binned approximation is
\begin{equation}
  W_1(p,q)\simeq
  \sum_{i=1}^{N_{\mathrm{bin}}-1}
  \left|\sum_{j=1}^{i}(p_j-q_j)\right|(y_{i+1}-y_i).
  \label{eq:wasserstein-binned}
\end{equation}
This is the one-dimensional analogue of the earth mover's distance used for
collider energy flows~\cite{Komiske:2019metric}. It measures the minimum probability-weighted transport distance along the
rapidity axis and is expressed in rapidity units. Thus, $\mathrm{TV}$
quantifies how much probability must be redistributed, whereas $W_1$
accounts for how far it must be moved. The values displayed in the plots
compare the baseline predictions; they characterize shape differences and
are not measures of statistical significance.

\subsection{Uncertainty treatment for \textsc{PYTHIA~8.306}}
\label{sec:uncertainties}
The uncertainty study for the \textsc{PYTHIA~8.306} predictions is performed
bin-by-bin on distributions normalized independently to unit area. The nominal
prediction uses the NNPDF2.3 LO set
(\texttt{NNPDF23\_lo\_as\_0130\_qed}), accessed through LHAPDF6,
and the Monash 2013 tune
(\texttt{Tune:pp = 14})~\cite{Skands:2014pea}. Starting from this baseline,
we vary the tune to CMS MonashStar (CUETP8M1-NNPDF2.3LO,
\texttt{Tune:pp = 18}) and the ATLAS A14 central tune with NNPDF2.3LO
(\texttt{Tune:pp = 21}), keeping the baseline PDF unchanged.
In a separate set of variations, we replace the baseline PDF with the
CT14 LO (\texttt{CT14lo}) or MMHT2014 LO
(\texttt{MMHT2014lo68cl}) set, also accessed through LHAPDF6, while
keeping the baseline Monash 2013 tune fixed.
The tune and PDF variations are thus evaluated separately relative to
the same nominal configuration.

Let $h_i^0$ be the nominal content of bin $i$, $\sigma_i^{\mathrm{stat}}$
its Monte Carlo statistical uncertainty, and $h_i^{(v)}$ the bin content for
an alternative sample. The asymmetric PDF envelope is defined as
\begin{align}
 \delta^{+}_{i,\mathrm{PDF}}
 &=\max\!\left(0,\,
 h_i^{\mathrm{CT14LO}}-h_i^0,\,
 h_i^{\mathrm{MMHT}}-h_i^0\right), \\
 \delta^{-}_{i,\mathrm{PDF}}
 &=\max\!\left(0,\,
 h_i^0-h_i^{\mathrm{CT14LO}},\,
 h_i^0-h_i^{\mathrm{MMHT}}\right),
 \label{eq:pdf-envelope}
\end{align}
and the tune envelope is constructed analogously,
\begin{align}
 \delta^{+}_{i,\mathrm{tune}}
 &=\max\!\left(0,\,
 h_i^{\mathrm{Tune\,18}}-h_i^0,\,
 h_i^{\mathrm{Tune\,21}}-h_i^0\right), \\
 \delta^{-}_{i,\mathrm{tune}}
 &=\max\!\left(0,\,
 h_i^0-h_i^{\mathrm{Tune\,18}},\,
 h_i^0-h_i^{\mathrm{Tune\,21}}\right).
 \label{eq:tune-envelope}
\end{align}
Only the shifts of the central values enter these two envelopes; statistical
errors from the alternative samples are not added. Assuming the nominal
statistical, PDF, and tune components to be independent, the total asymmetric
uncertainty is
\begin{equation}
 \delta^{\pm}_{i,\mathrm{tot}}
 =
 \sqrt{
   \left(\sigma_i^{\mathrm{stat}}\right)^2
   +\left(\delta^{\pm}_{i,\mathrm{PDF}}\right)^2
   +\left(\delta^{\pm}_{i,\mathrm{tune}}\right)^2
 }.
 \label{eq:total-envelope}
\end{equation}
Since each input histogram is normalized before the comparison, the bands quantify
shape variations rather than changes in the overall rate.

\subsection{Uncertainty treatment for \textsc{BFKLex}}
\label{sec:bfklex-uncertainties}
\providecommand{\BFKLLambdaFigDir}{figures/lambda_BFKLex_comparison} 

The BFKLex predictions are obtained at leading-logarithmic (LL) accuracy.
The nominal renormalization and factorization scales are
$\mu_R=\mu_F=\mu_0=\sqrt{k_a k_b}$, with $k_a$ and $k_b$ denoting the
transverse-momentum magnitudes of the two partonic endpoints of the BFKL
ladder. We distinguish the dependence on perturbative scales and PDF-set
inputs from the numerical stability with respect to the infrared resolution
parameter $\lambda_{\mathrm{IR}}$. The uncertainty bands discussed below
refer to distributions independently normalized to unit area.

To assess the perturbative-scale dependence, we consider the nominal point
$(\mu_R/\mu_0,\mu_F/\mu_0)=(1,1)$ and the six variations
$(1/2,1)$, $(2,1)$, $(1,1/2)$, $(1,2)$, $(1/2,1/2)$, and $(2,2)$.
The two extreme anti-correlated combinations are not included. We group
the six alternatives into renormalization-scale variations ($R$),
factorization-scale variations ($F$), and simultaneous variations of both
scales ($RF$). For each group $s\in\{R,F,RF\}$, we form a one-sided,
bin-by-bin envelope about the nominal unit-area density $h_i^{(0)}$:
\begin{equation}
\begin{aligned}
 \delta_{i,s}^{+} &=
 \max\!\left(0,\max_{v\in\mathcal{V}_s}
                  [h_i^{(v)}-h_i^{(0)}]\right),\\
 \delta_{i,s}^{-} &=
 \max\!\left(0,\max_{v\in\mathcal{V}_s}
                  [h_i^{(0)}-h_i^{(v)}]\right).
\end{aligned}
\label{eq:bfklex-grouped-envelopes}
\end{equation}
Here $h_i^{(v)}$ denotes the independently normalized result for variation
$v$. We adopt a deliberately conservative prescription in which the three
grouped scale envelopes are combined in quadrature,
\begin{equation}
 \delta_{i,\mathrm{scale}}^{\pm}=
 \sqrt{(\delta_{i,R}^{\pm})^2+
       (\delta_{i,F}^{\pm})^2+
       (\delta_{i,RF}^{\pm})^2}\,.
\label{eq:bfklex-conservative-scale}
\end{equation}
This prescription gives a band at least as wide, bin by bin, as taking a
single envelope over all six alternative scale choices. The simultaneous
variations are retained to allow for a correlated scale response, while the
quadrature combination is used as a conservative convention; it should not
be interpreted as a statistical independence statement about the scale
variations.

The nominal PDF set is NNPDF2.3 LO
(\texttt{NNPDF23\_lo\_as\_0130\_qed}), with CT14LO and
MMHT2014lo68cl as alternative PDF-set inputs. The PDF-set/input envelope
is constructed from their one-sided shifts about the nominal, again using
independently unit-normalized distributions. For each set we use the
corresponding $\alpha_s$ implementation and running supplied with that
PDF input, rather than fixing $\alpha_s$ to the nominal set's prescription.
The resulting variation therefore probes the combined change of PDF input
and its associated $\alpha_s$ treatment, and is not a fixed-$\alpha_s$
PDF eigenvector or replica uncertainty. We combine this envelope with the
scale prescription above and the nominal Monte Carlo statistical error:
\begin{equation}
 \delta_{i,\mathrm{tot}}^{\pm}=
 \sqrt{(\sigma_{i,\mathrm{MC}})^2+
       (\delta_{i,\mathrm{PDF\,input}}^{\pm})^2+
       (\delta_{i,\mathrm{scale}}^{\pm})^2}\,.
\label{eq:bfklex-combined-band}
\end{equation}
The statistical contribution $\sigma_{i,\mathrm{MC}}$ is obtained from
the nominal weighted histogram using the sum of squared event weights,
rescaled by the nominal histogram's unit-area normalization. Statistical
errors of alternative samples are not included separately in the
central-value envelopes. The final band is a pragmatic measure of the
variations considered here, not a confidence interval or a complete
estimate of corrections beyond LL accuracy.

As a separate numerical-stability test, we vary the infrared resolution
parameter over $\lambda_{\mathrm{IR}}=0.5$, $1$, and $2\,\mathrm{GeV}$,
with $1\,\mathrm{GeV}$ nominal. This variation does not contribute to the
theory uncertainty bands defined above. The generated BFKL evolution is
truncated at a maximum of 20 emissions, in addition to the two partonic
endpoints of the ladder. To test convergence of the emission sum, the
dominant low-transverse-momentum region was explicitly extended to a
maximum of 25 emissions (25 emitted partons plus the two endpoints).
The contributions from emissions 21--25 were included in the final
nominal and varied predictions; outside that region the original
20-emission truncation was retained. The observed convergence of the
high-multiplicity tail and the small corrections to the predictions
support the adequacy of this truncation.

\begin{figure}[!t]
  \centering
    \includegraphics[width=0.613525\textwidth,pagebox=mediabox,trim=250bp 45bp 25bp 45bp,clip]{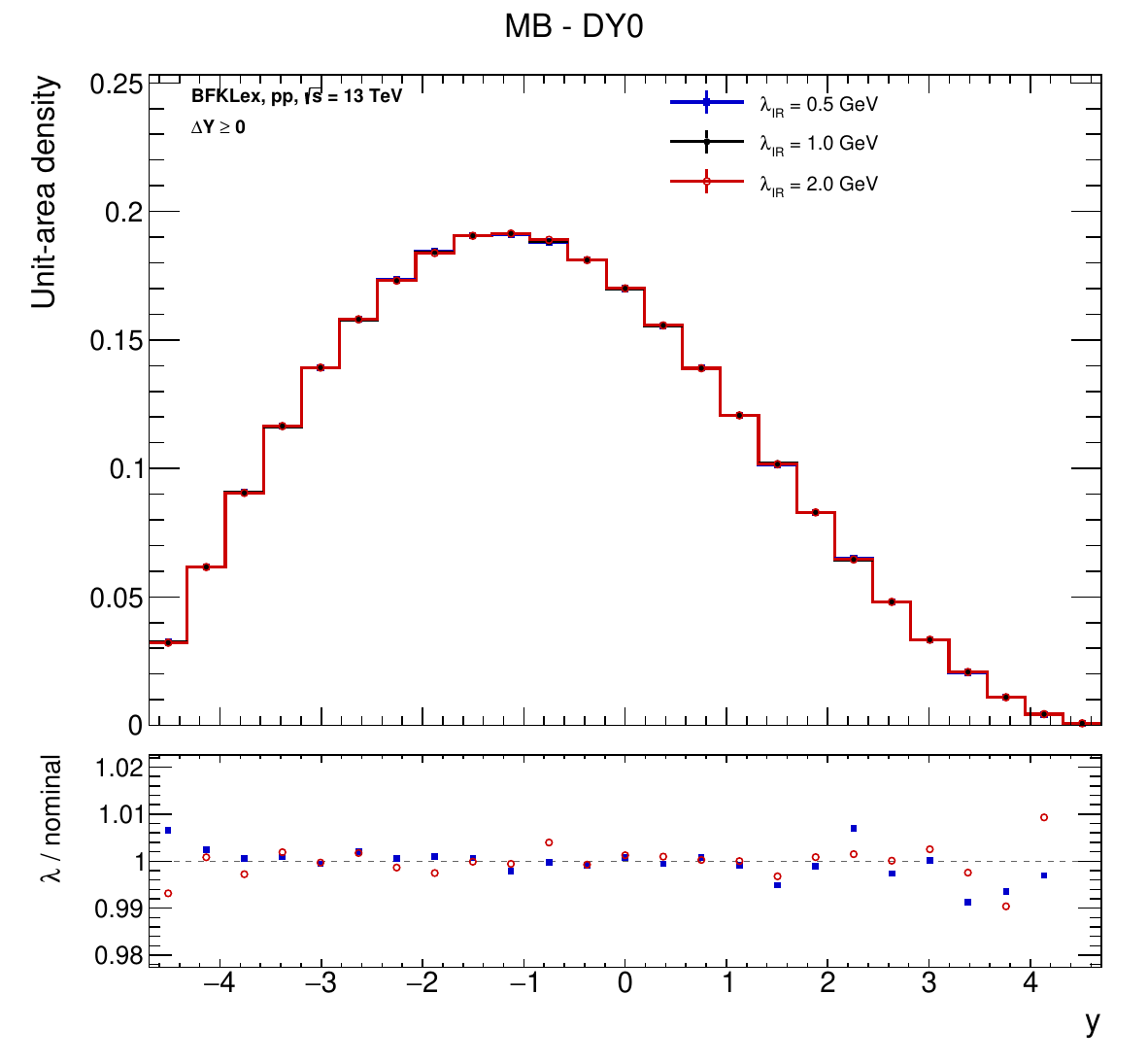}
    \caption{Infrared-resolution stability of the unit-normalized most-backward (MB) rapidity distribution in BFKLex, for $\Delta Y_{\mathrm{MF,MB}}\geq0$. The curves correspond to $\lambda_{\mathrm{IR}}=0.5$, $1$, and $2\,\mathrm{GeV}$. The lower panel shows the ratios to the nominal $\lambda_{\mathrm{IR}}=1\,\mathrm{GeV}$ prediction.}
    \label{fig:bfklex-lambda-mb}
\end{figure}

\begin{figure}[!t]
  \centering
    \includegraphics[width=0.613525\textwidth,pagebox=mediabox,trim=250bp 45bp 25bp 45bp,clip]{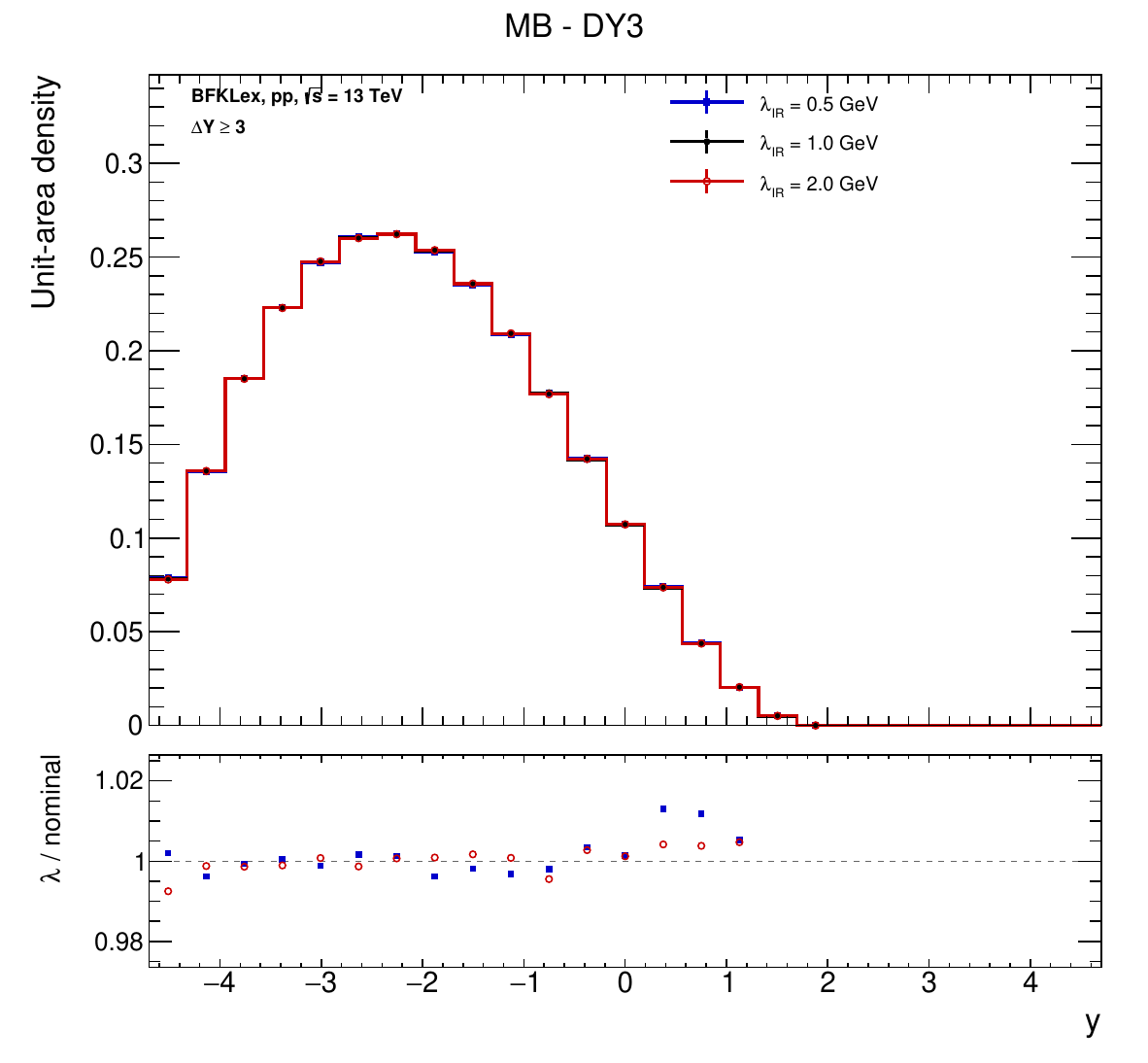}
    \caption{Infrared-resolution stability of the unit-normalized most-backward (MB) rapidity distribution in BFKLex, for $\Delta Y_{\mathrm{MF,MB}}\geq3$. The curves correspond to $\lambda_{\mathrm{IR}}=0.5$, $1$, and $2\,\mathrm{GeV}$. The lower panel shows the ratios to the nominal $\lambda_{\mathrm{IR}}=1\,\mathrm{GeV}$ prediction.}
    \label{fig:bfklex-lambda-mb-dy3}
\end{figure}

\begin{figure}[!t]
  \centering
    \includegraphics[width=0.613525\textwidth,pagebox=mediabox,trim=250bp 45bp 25bp 45bp,clip]{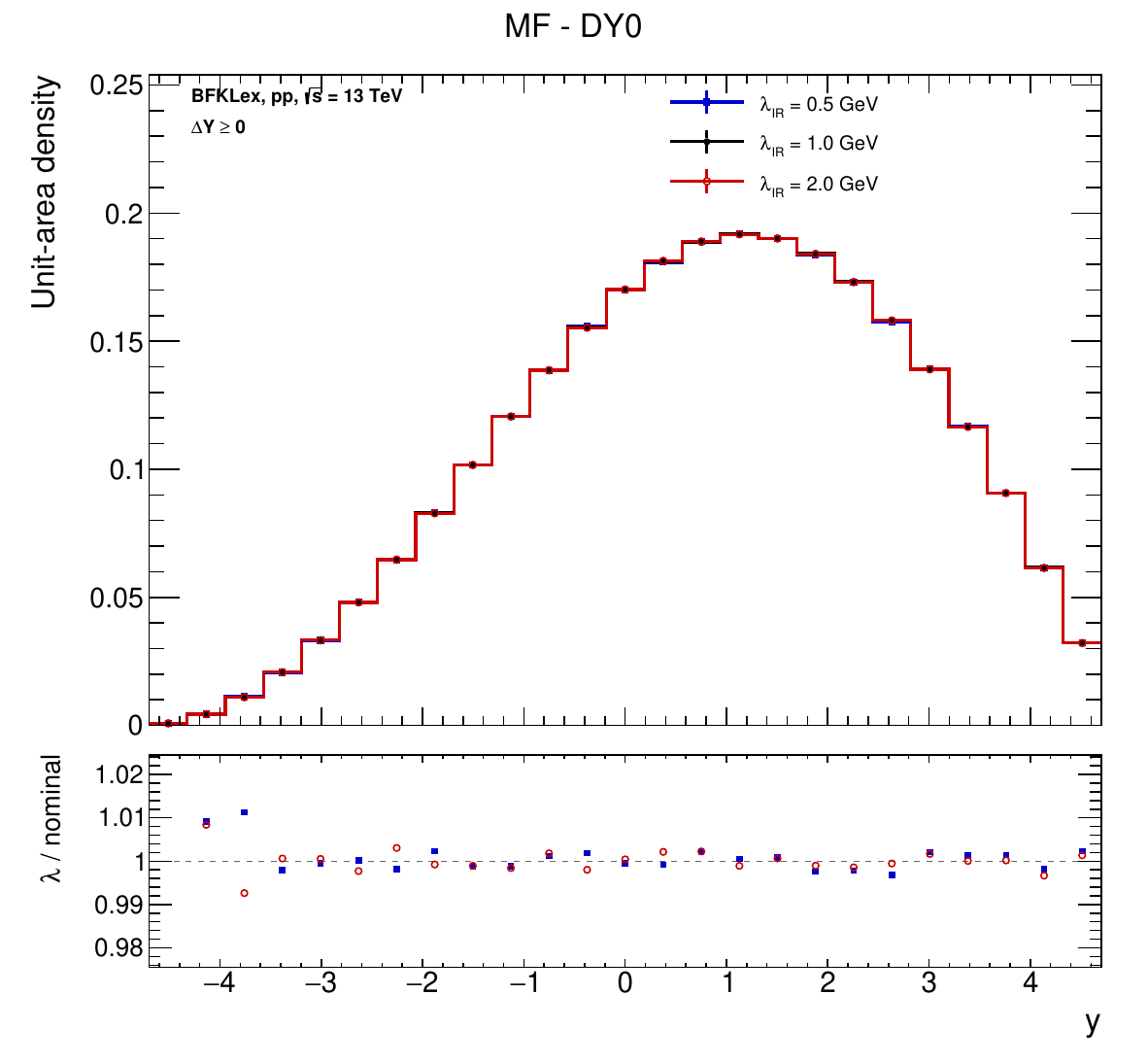}
    \caption{As in Fig.~\ref{fig:bfklex-lambda-mb}, for the most-forward (MF) rapidity distribution with $\Delta Y_{\mathrm{MF,MB}}\geq0$.}
    \label{fig:bfklex-lambda-mf}
\end{figure}

\begin{figure}[!t]
  \centering
    \includegraphics[width=0.613525\textwidth,pagebox=mediabox,trim=250bp 45bp 25bp 45bp,clip]{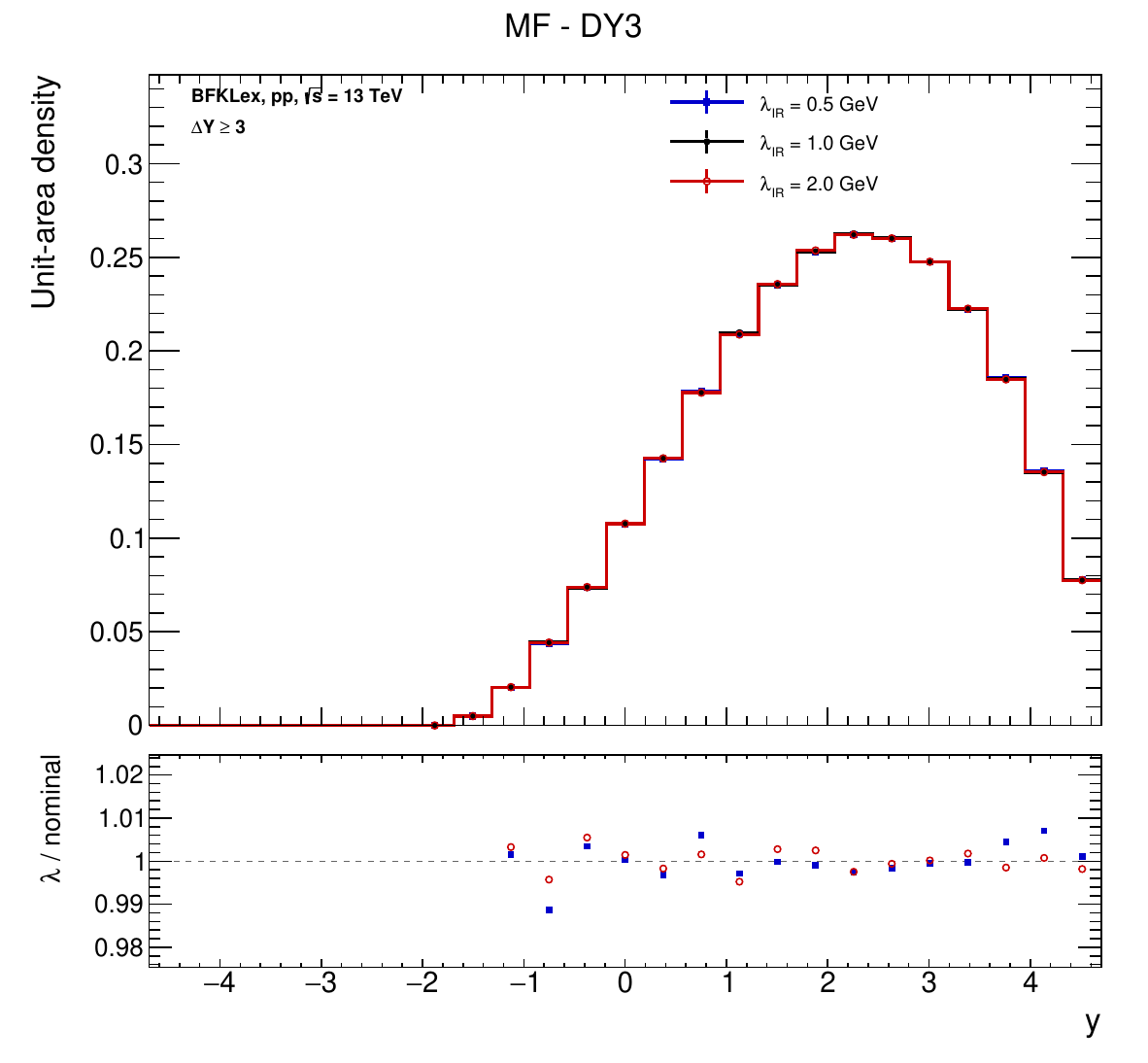}
    \caption{As in Fig.~\ref{fig:bfklex-lambda-mb}, for the most-forward (MF) rapidity distribution with $\Delta Y_{\mathrm{MF,MB}}\geq3$.}
    \label{fig:bfklex-lambda-mf-dy3}
\end{figure}

\begin{figure}[!t]
  \centering
    \includegraphics[width=0.613525\textwidth,pagebox=mediabox,trim=250bp 45bp 25bp 45bp,clip]{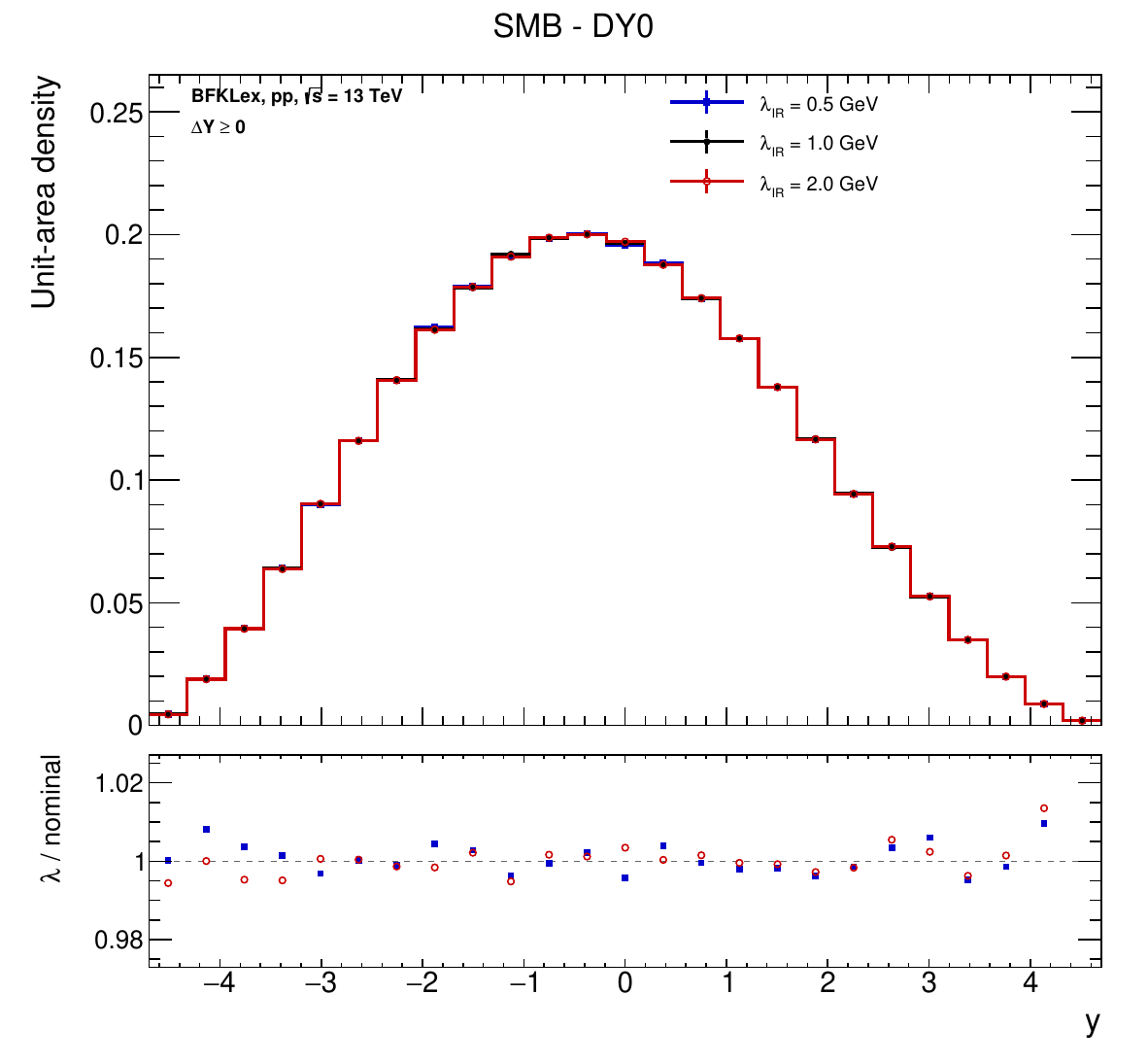}
    \caption{As in Fig.~\ref{fig:bfklex-lambda-mb}, for the second-most-backward (SMB) rapidity distribution with $\Delta Y_{\mathrm{MF,MB}}\geq0$.}
    \label{fig:bfklex-lambda-smb}
\end{figure}

\begin{figure}[!t]
  \centering
    \includegraphics[width=0.613525\textwidth,pagebox=mediabox,trim=250bp 45bp 25bp 45bp,clip]{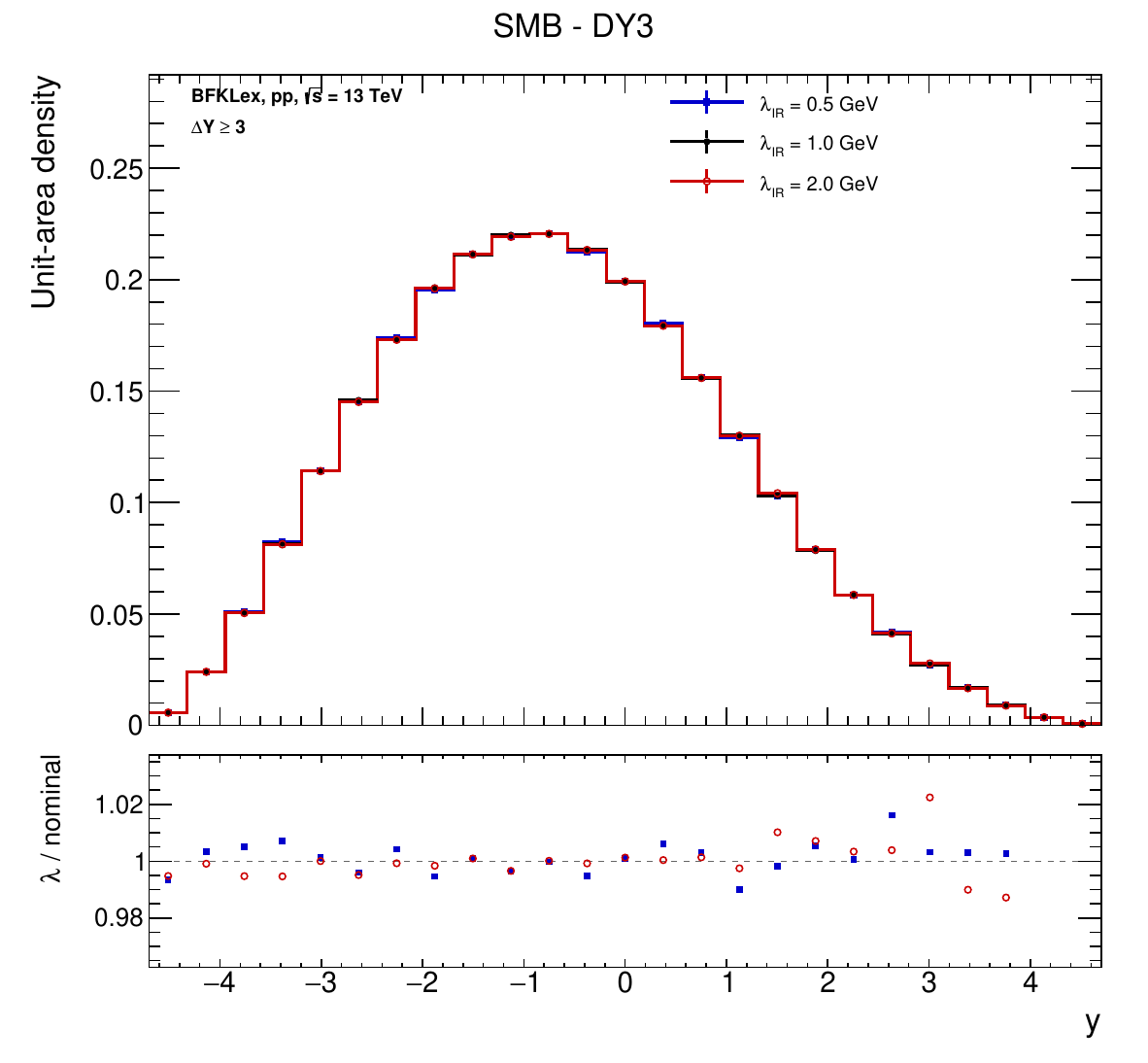}
    \caption{As in Fig.~\ref{fig:bfklex-lambda-mb}, for the second-most-backward (SMB) rapidity distribution with $\Delta Y_{\mathrm{MF,MB}}\geq3$.}
    \label{fig:bfklex-lambda-smb-dy3}
\end{figure}

\begin{figure}[!t]
  \centering
    \includegraphics[width=0.613525\textwidth,pagebox=mediabox,trim=250bp 45bp 25bp 45bp,clip]{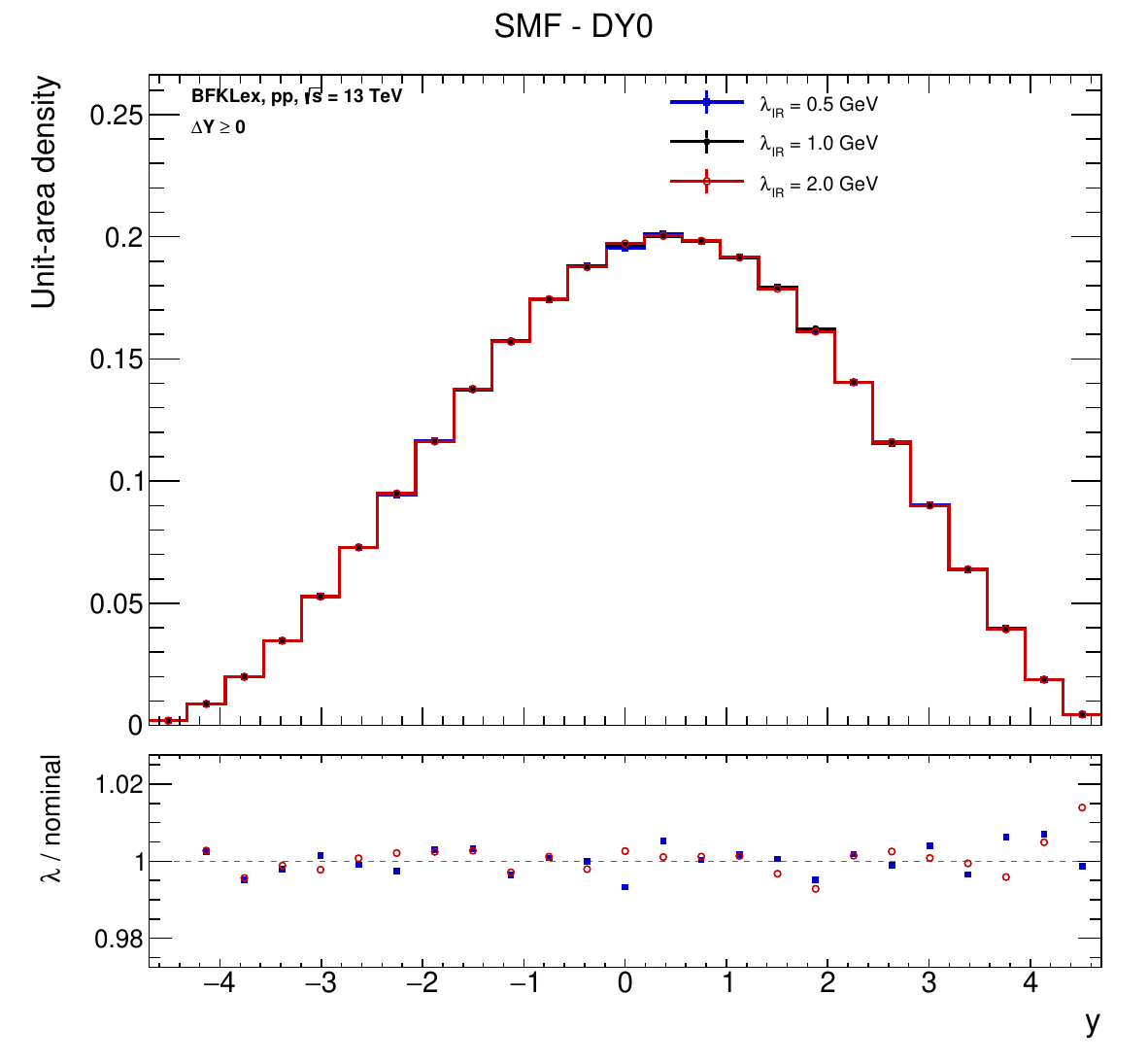}
    \caption{As in Fig.~\ref{fig:bfklex-lambda-mb}, for the second-most-forward (SMF) rapidity distribution with $\Delta Y_{\mathrm{MF,MB}}\geq0$.}
    \label{fig:bfklex-lambda-smf}
\end{figure}

\begin{figure}[!t]
  \centering
    \includegraphics[width=0.613525\textwidth,pagebox=mediabox,trim=250bp 45bp 25bp 45bp,clip]{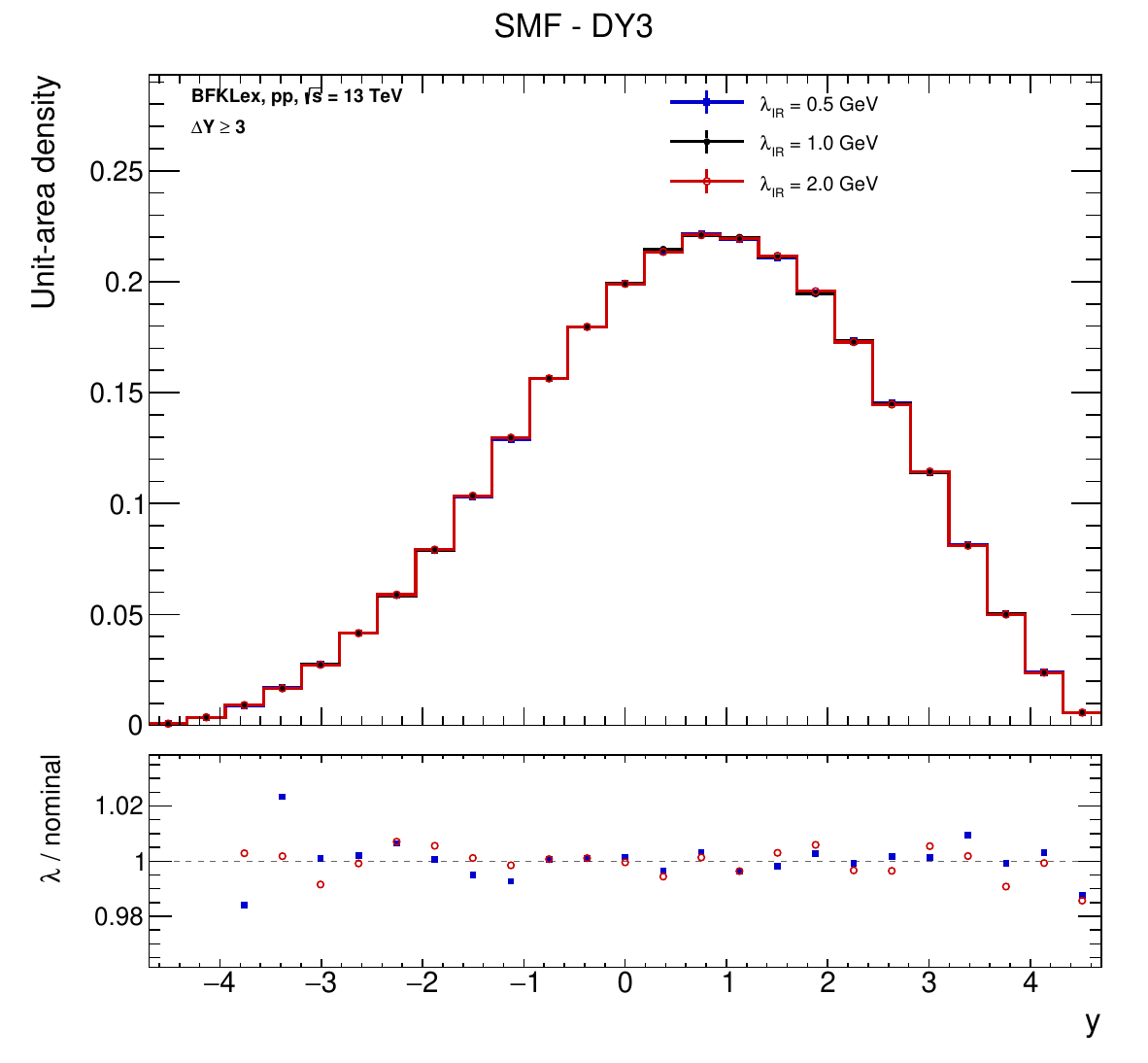}
    \caption{As in Fig.~\ref{fig:bfklex-lambda-mb}, for the second-most-forward (SMF) rapidity distribution with $\Delta Y_{\mathrm{MF,MB}}\geq3$.}
    \label{fig:bfklex-lambda-smf-dy3}
\end{figure}

\begin{figure}[!t]
  \centering
    \includegraphics[width=0.613525\textwidth,pagebox=mediabox,trim=250bp 45bp 25bp 45bp,clip]{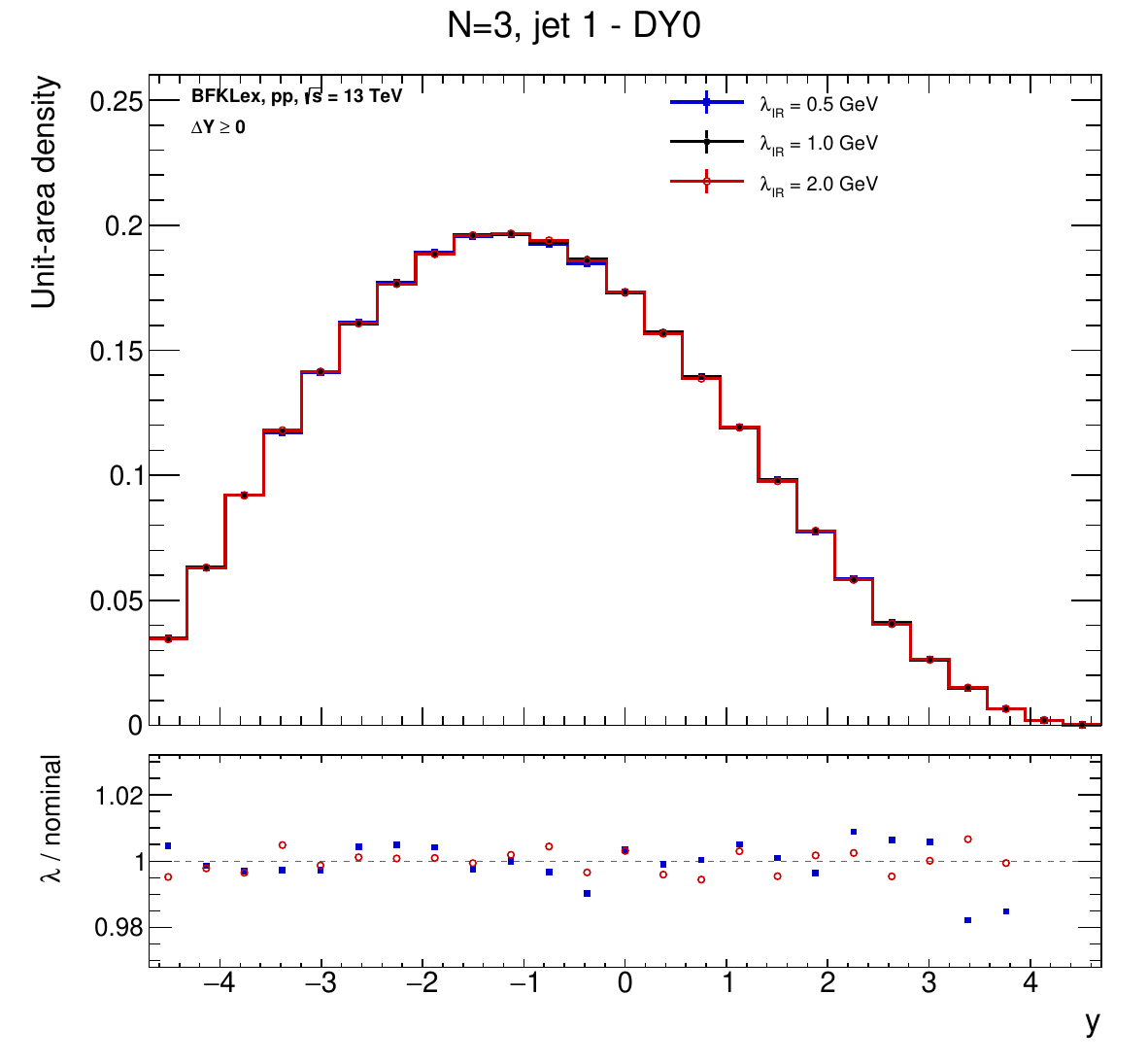}
    \caption{As in Fig.~\ref{fig:bfklex-lambda-mb}, for the most-backward jet in events with exactly three accepted jets ($N=3$, jet~1) with $\Delta Y_{\mathrm{MF,MB}}\geq0$.}
    \label{fig:bfklex-lambda-eta3-0}
\end{figure}

\begin{figure}[!t]
  \centering
    \includegraphics[width=0.613525\textwidth,pagebox=mediabox,trim=250bp 45bp 25bp 45bp,clip]{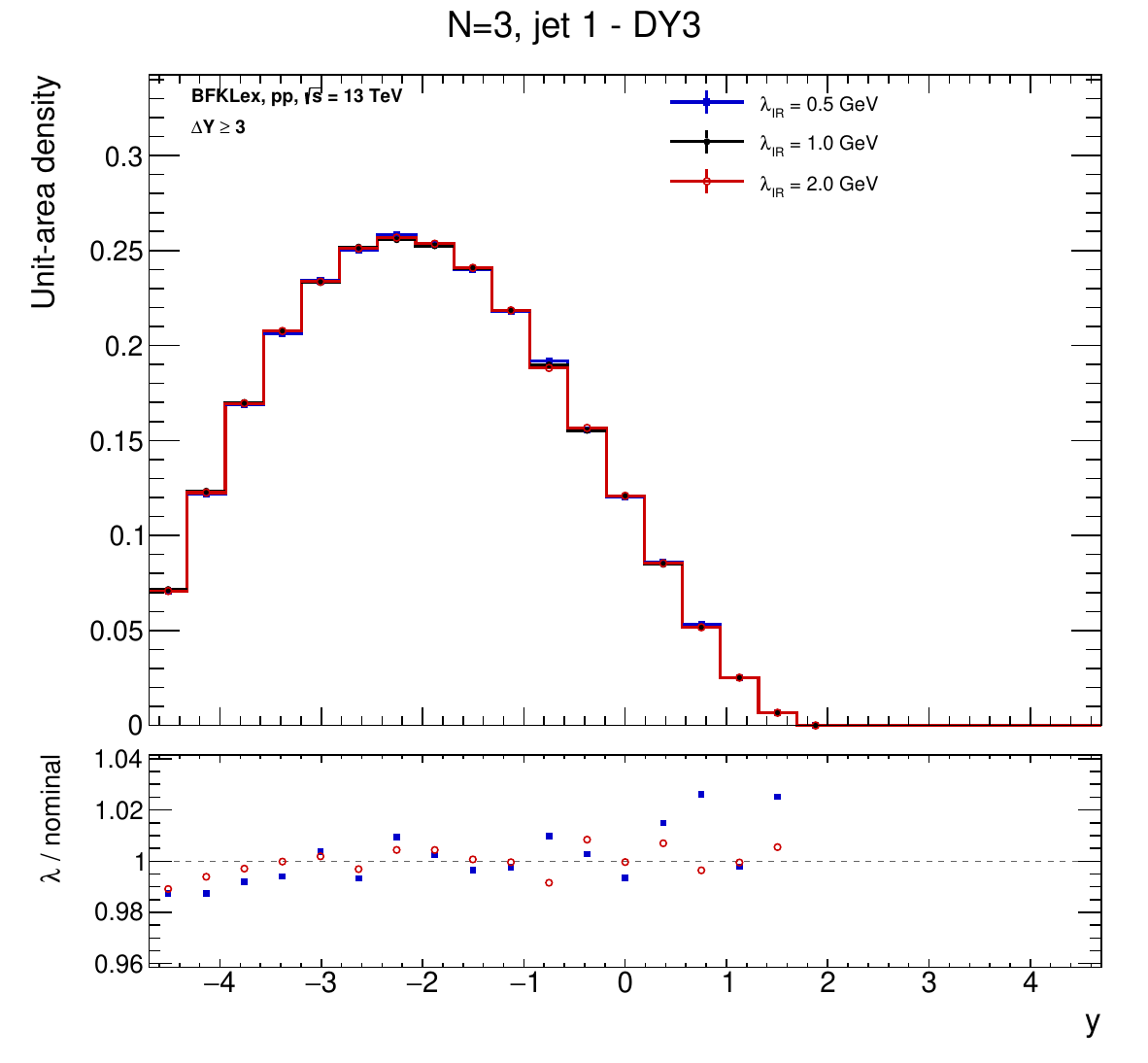}
    \caption{As in Fig.~\ref{fig:bfklex-lambda-mb}, for the most-backward jet in events with exactly three accepted jets ($N=3$, jet~1) with $\Delta Y_{\mathrm{MF,MB}}\geq3$.}
    \label{fig:bfklex-lambda-eta3-0-dy3}
\end{figure}

\begin{figure}[!t]
  \centering
    \includegraphics[width=0.613525\textwidth,pagebox=mediabox,trim=250bp 45bp 25bp 45bp,clip]{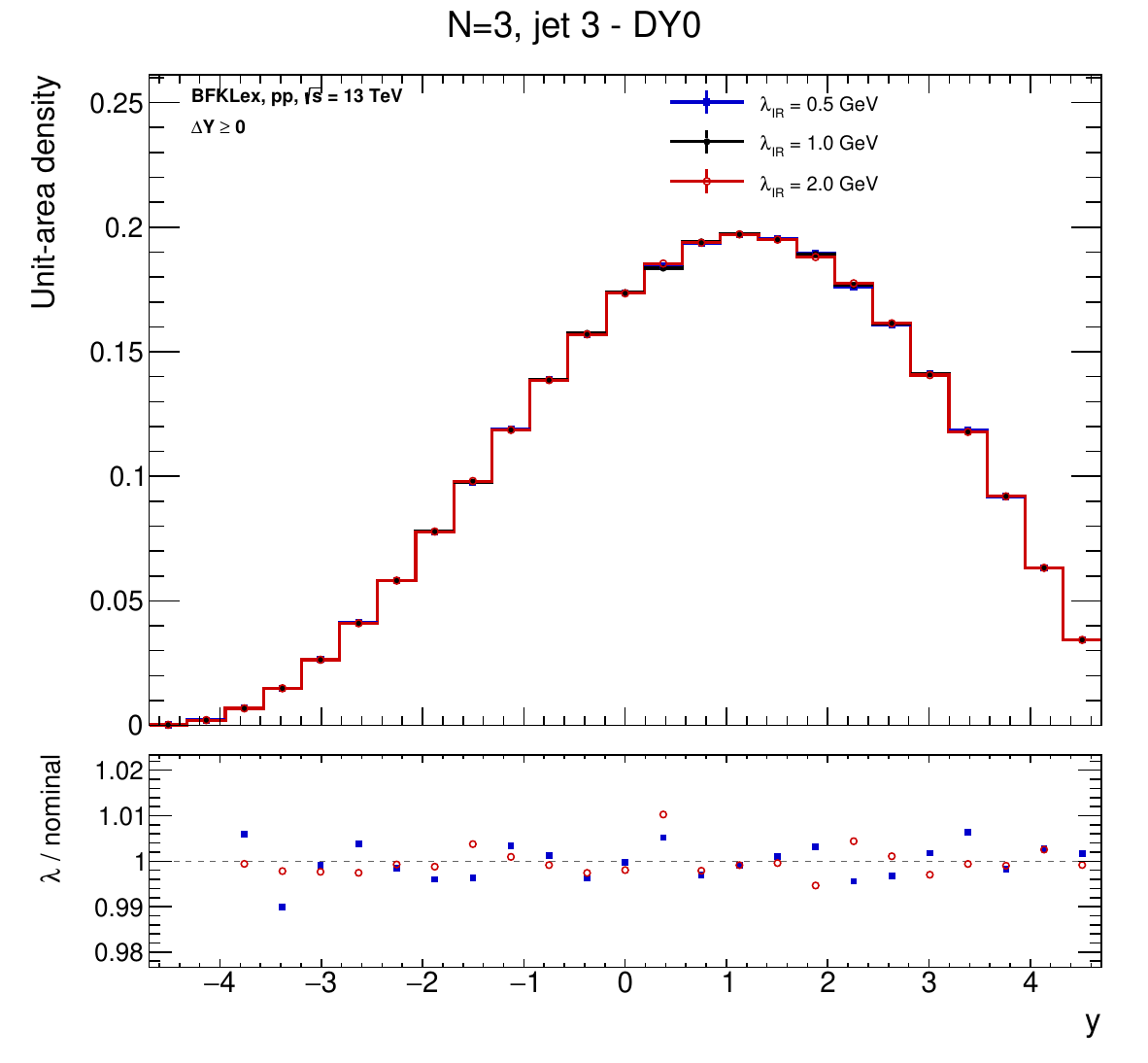}
    \caption{As in Fig.~\ref{fig:bfklex-lambda-mb}, for the most-forward jet in events with exactly three accepted jets ($N=3$, jet~3) with $\Delta Y_{\mathrm{MF,MB}}\geq0$.}
    \label{fig:bfklex-lambda-eta3-2}
\end{figure}

\begin{figure}[!t]
  \centering
    \includegraphics[width=0.613525\textwidth,pagebox=mediabox,trim=250bp 45bp 25bp 45bp,clip]{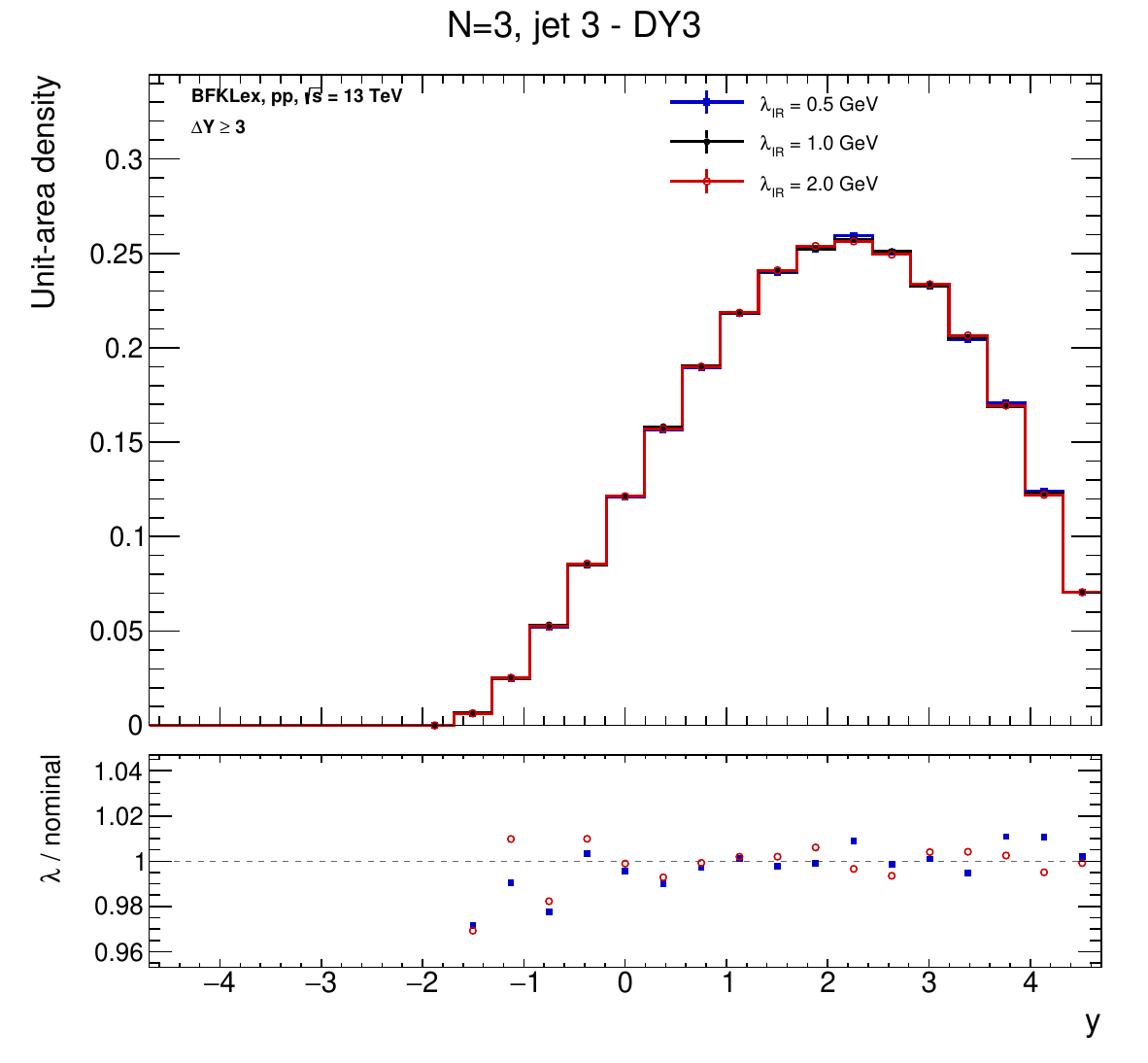}
    \caption{As in Fig.~\ref{fig:bfklex-lambda-mb}, for the most-forward jet in events with exactly three accepted jets ($N=3$, jet~3) with $\Delta Y_{\mathrm{MF,MB}}\geq3$.}
    \label{fig:bfklex-lambda-eta3-2-dy3}
\end{figure}

\begin{figure}[!t]
  \centering
    \includegraphics[width=0.613525\textwidth,pagebox=mediabox,trim=250bp 45bp 25bp 45bp,clip]{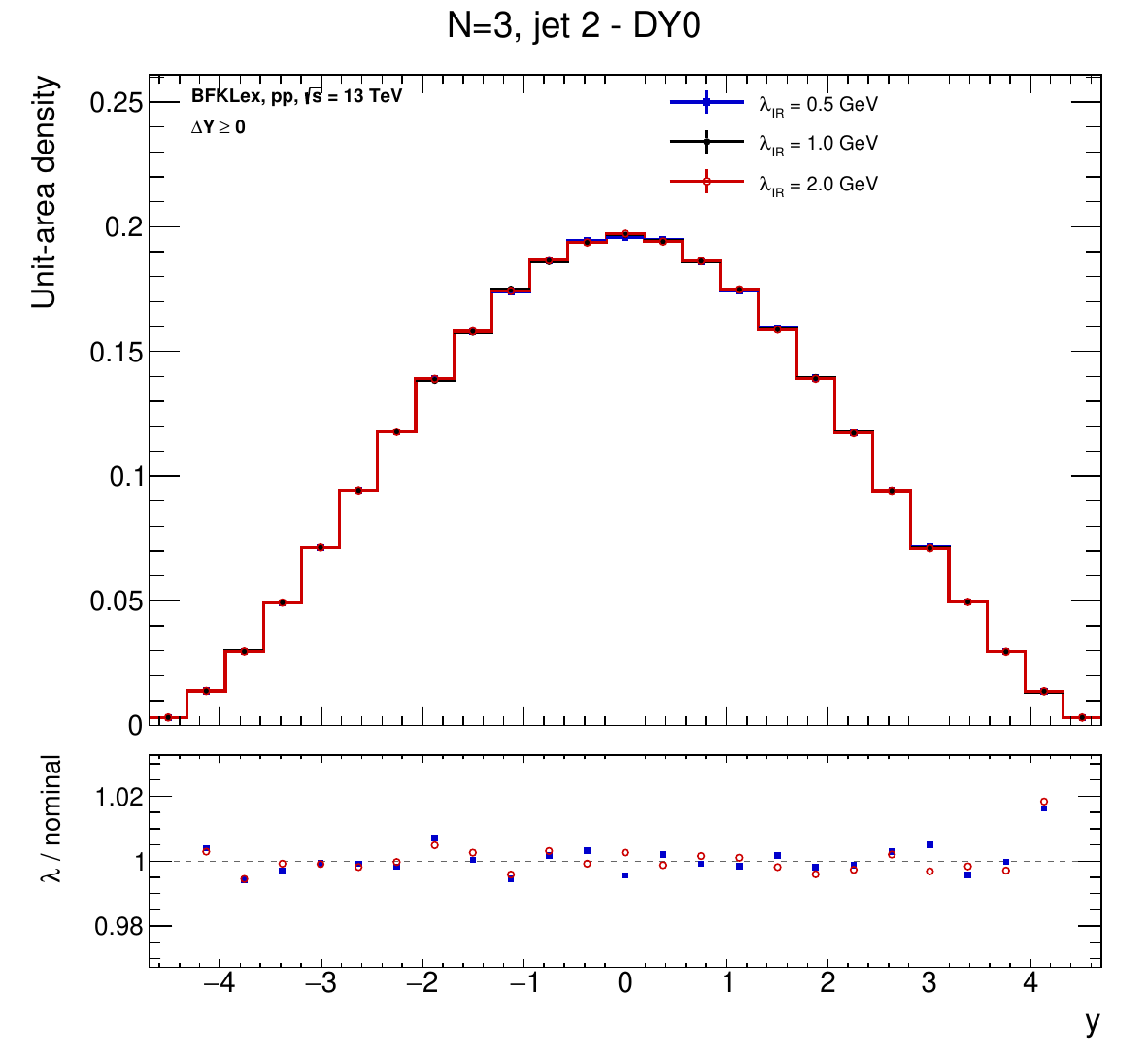}
    \caption{As in Fig.~\ref{fig:bfklex-lambda-mb}, for the middle jet in events with exactly three accepted jets ($N=3$, jet~2) with $\Delta Y_{\mathrm{MF,MB}}\geq0$.}
    \label{fig:bfklex-lambda-eta3-1}
\end{figure}

\begin{figure}[!t]
  \centering
    \includegraphics[width=0.613525\textwidth,pagebox=mediabox,trim=250bp 45bp 25bp 45bp,clip]{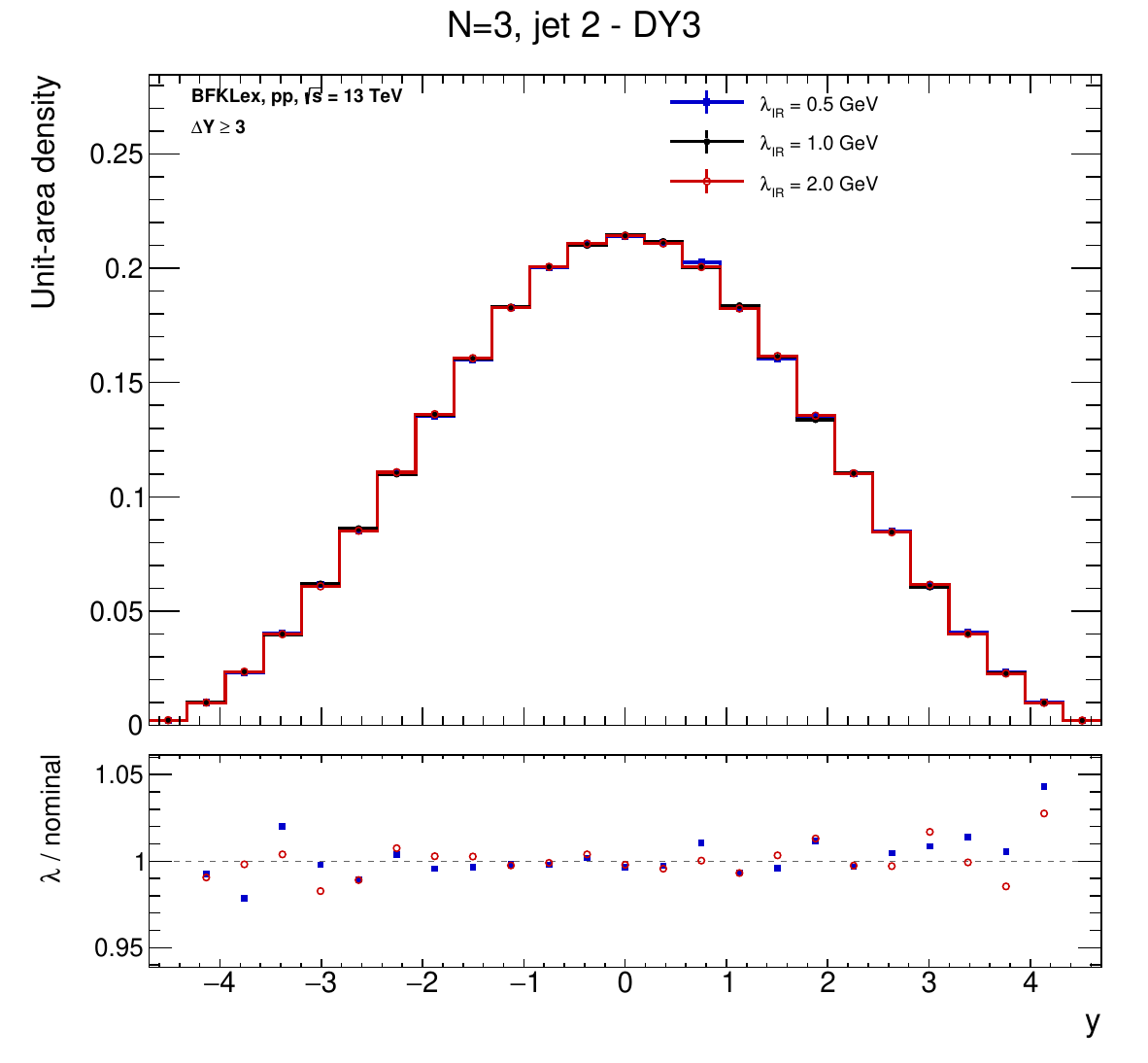}
    \caption{As in Fig.~\ref{fig:bfklex-lambda-mb}, for the middle jet in events with exactly three accepted jets ($N=3$, jet~2) with $\Delta Y_{\mathrm{MF,MB}}\geq3$.}
    \label{fig:bfklex-lambda-eta3-1-dy3}
\end{figure}

\paragraph{Infrared-resolution stability.}
Figures~\ref{fig:bfklex-lambda-mb}--\ref{fig:bfklex-lambda-eta3-1-dy3} present the
infrared-resolution check for the rapidity ranks. For every observable,
the $\Delta Y_{\mathrm{MF,MB}}\geq0$ and $\Delta Y_{\mathrm{MF,MB}}\geq3$ selections are shown in successive figures, so that the effect of the additional rapidity-separation requirement
can be assessed directly. The three curves in each panel correspond to
$\lambda_{\mathrm{IR}}=0.5$, $1$, and $2\,\mathrm{GeV}$, normalized
independently to unit area; the lower panels show their ratios to the
$1\,\mathrm{GeV}$ prediction. The distributions remain remarkably stable
under this variation, for the inclusive-extremal and second-extremal ranks
as well as for the three ranks at fixed jet multiplicity $N=3$.

For completeness, the accepted rates change by $+0.08\%$ and $-0.27\%$
in $\Delta Y_{\mathrm{MF,MB}}\geq0$ when varying $\lambda_{\mathrm{IR}}$ to $0.5$ and $2\,\mathrm{GeV}$,
respectively, relative to the nominal result. In $\Delta Y_{\mathrm{MF,MB}}\geq3$ the corresponding
changes are $+0.35\%$ and $-0.40\%$. The unit-normalized shapes are
considerably less sensitive: the largest total-variation distances from
the nominal shapes over the seven observables and two cutoff variations
are $\mathrm{TV}=0.00181$ for $\Delta Y_{\mathrm{MF,MB}}\geq0$ and $0.00337$ for $\Delta Y_{\mathrm{MF,MB}}\geq3$.
The infrared-cutoff variations are therefore retained as a numerical
stability check, with $\lambda_{\mathrm{IR}}=1\,\mathrm{GeV}$ used for
the subsequent phenomenological comparisons.

\paragraph{Perturbative-scale and PDF-set sensitivity.}
The perturbative-scale dependence is substantially greater than that
associated with the infrared cutoff. At the level of the accepted rates,
changing $\mu_R$ alone by factors of $1/2$ and $2$ changes the $\Delta Y_{\mathrm{MF,MB}}\geq0$
prediction by approximately $+59\%$ and $-32\%$, respectively; the
corresponding shifts for $\Delta Y_{\mathrm{MF,MB}}\geq3$ are about $+76\%$ and $-37\%$.
Factorization-scale variations alone give much smaller changes in the
integrated rates. Unit-area normalization removes much of this overall
scale response but does not eliminate shape changes. The second-extremal
ranks, especially with the $\Delta Y_{\mathrm{MF,MB}}\geq3$ separation requirement, are more sensitive
to variations involving $\mu_R$, whereas PDF-set/input and
factorization-scale changes are relatively more important for the outer
jets at fixed $N=3$. This observable dependence motivates keeping the
scale and PDF-set/input contributions explicit in the uncertainty
assessment.

\vspace{-\baselineskip}
\section{Results}

The distributions in this section compare \textsc{PYTHIA~8.306} and
\textsc{BFKLex} predictions at $\sqrt{s}=13$~TeV with the selection in
Eq.~\eqref{eq:asymmetric-boundary-cuts}. Each distribution is normalized to
unit area. Blue and red baseline curves denote \textsc{PYTHIA~8.306} and
\textsc{BFKLex}, respectively, with statistical error bars. The shaded bands
show the uncertainty components and their combinations indicated in the
legends. The lower panels show the ratio of the
\textsc{BFKLex} to the \textsc{PYTHIA~8.306} baseline prediction.

\subsection{No additional gap requirement}

Figures~\ref{fig:N3-Y0}--\ref{fig:SMFSMB-Y0} show the rank-ordered spectra for
$\Delta Y_{\mathrm{MF,MB}}\geq 0$. The three-jet distributions resolve the
three rapidity ranks separately, whereas the MB/MF and SMB/SMF projections
include all accepted jet multiplicities.

For fixed multiplicity $N=3$, Fig.~\ref{fig:N3-Y0} shows systematic
shape differences between the two generators. The \textsc{PYTHIA~8.306}
distributions have more pronounced maxima, whereas \textsc{BFKLex}
assigns more probability to rapidities away from these maxima. The outer
ranks exhibit both a reduced peak height and a redistribution of
probability along the rapidity axis. Their total-variation distances are
approximately $0.12$, with Wasserstein distances of approximately $0.46$.

\begin{figure}[!htbp]
\centering
\begin{subfigure}[t]{0.70\textwidth}
    \centering
    \includegraphics[width=\linewidth,trim=0 0 0 23,clip]{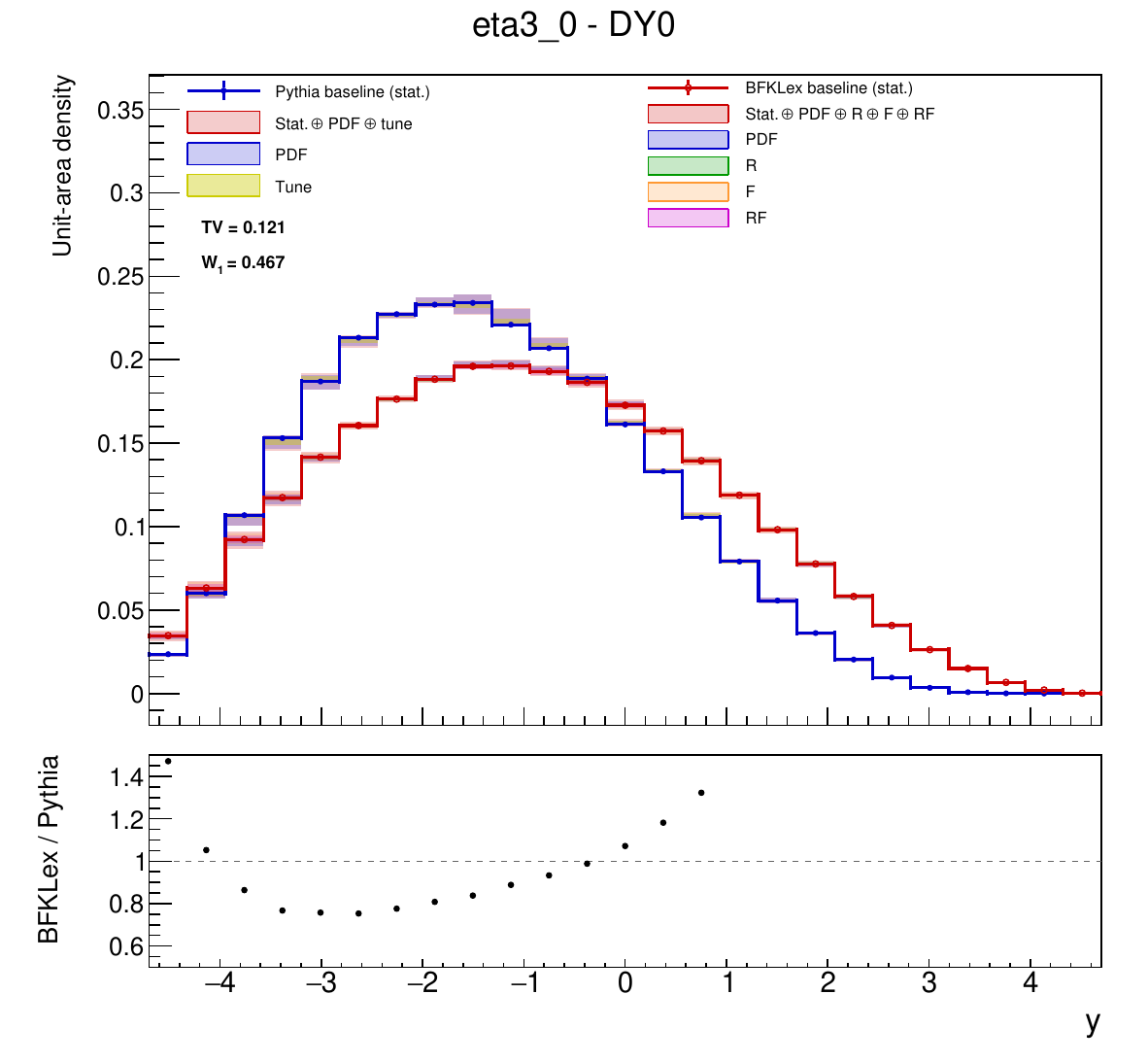}
    \caption{Jet1 (most backward).}
  \end{subfigure}
\caption{\textbf{Fixed multiplicity $N=3$ and
  $\Delta Y_{\mathrm{MF,MB}}\geq0$.} Normalized rapidity spectra for the three ordered jets, comparing \textsc{PYTHIA~8.306} and \textsc{BFKLex}. Bands denote the uncertainty components specified in the legends; lower panels show the \textsc{BFKLex}/\textsc{PYTHIA~8.306} baseline ratio.}
  \label{fig:N3-Y0}
\end{figure}

\begin{figure}[!htbp]
\ContinuedFloat
\centering
\begin{subfigure}[t]{0.70\textwidth}
    \centering
    \includegraphics[width=\linewidth,trim=0 0 0 23,clip]{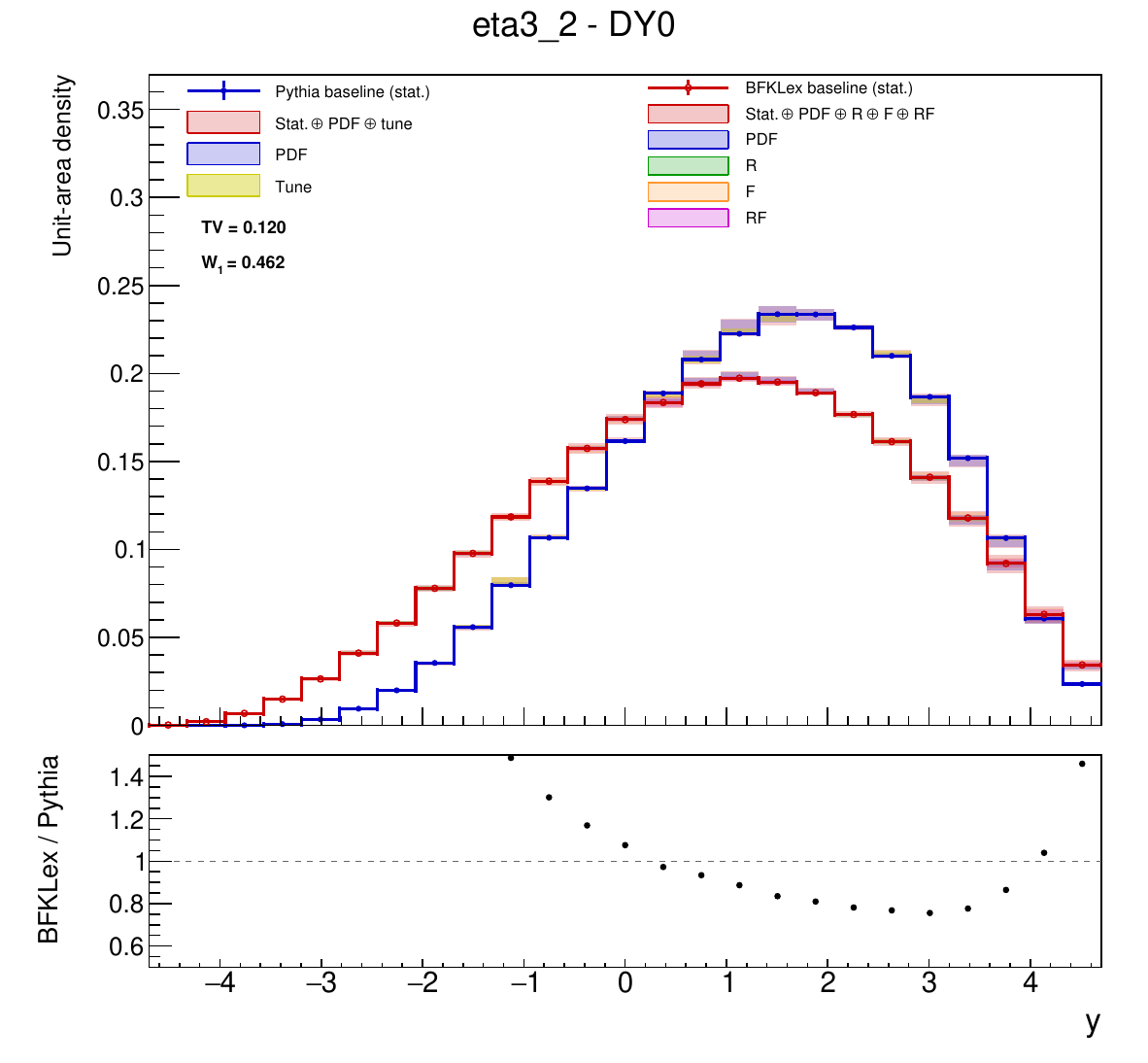}
    \caption{Jet3 (most forward).}
  \end{subfigure}
\captionsetup{labelformat=continuedfigure}
\caption[]{Corresponding forward-rank distribution for the same selection.}
\end{figure}

For the central rank, both distributions are approximately symmetric about
zero: the main difference is their width, with a broader \textsc{BFKLex}
spectrum. The separation is smaller than for the outer ranks, with
$\mathrm{TV}=0.0916$ and $W_1=0.237$.

\begin{figure}[!htbp]
  \ContinuedFloat
  \centering
\begin{subfigure}[t]{0.70\textwidth}
    \centering
    \includegraphics[width=\linewidth,trim=0 0 0 23,clip]{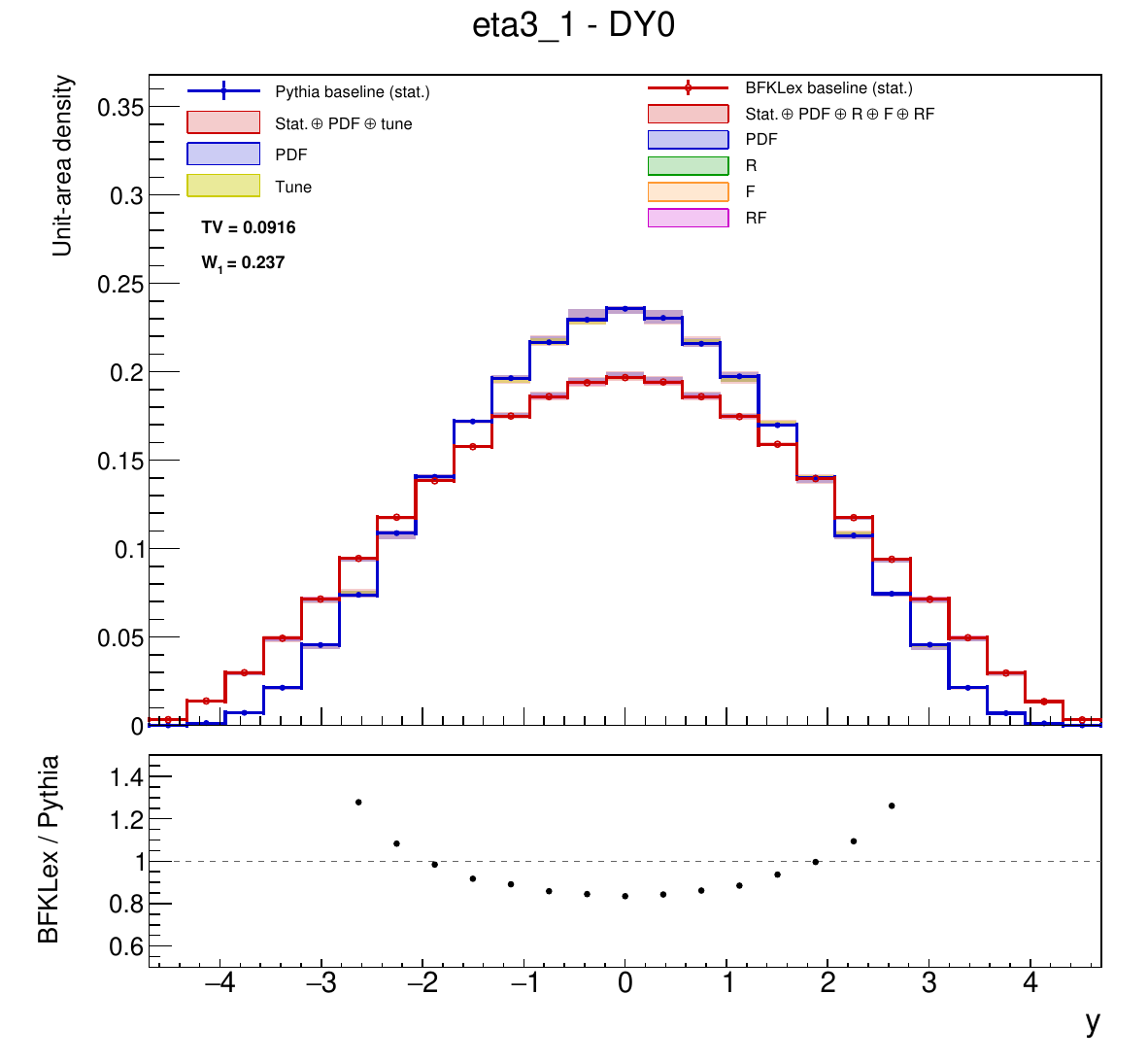}
    \caption{Jet2 (central rank).}
  \end{subfigure}
  \captionsetup{labelformat=continuedfigure}
  \caption[]{Central-rank rapidity spectrum for the same fixed-multiplicity selection.}
\end{figure}

The observables defined across multiplicities show a clear dependence on
rank. The MB and MF spectra in Fig.~\ref{fig:MBMF-Y0} are relatively close,
with $\mathrm{TV}\simeq0.055$ and $W_1\simeq0.20$.

\begin{figure}[!htbp]
\centering
\begin{subfigure}[t]{0.70\textwidth}
    \centering
    \includegraphics[width=\linewidth,trim=0 0 0 23,clip]{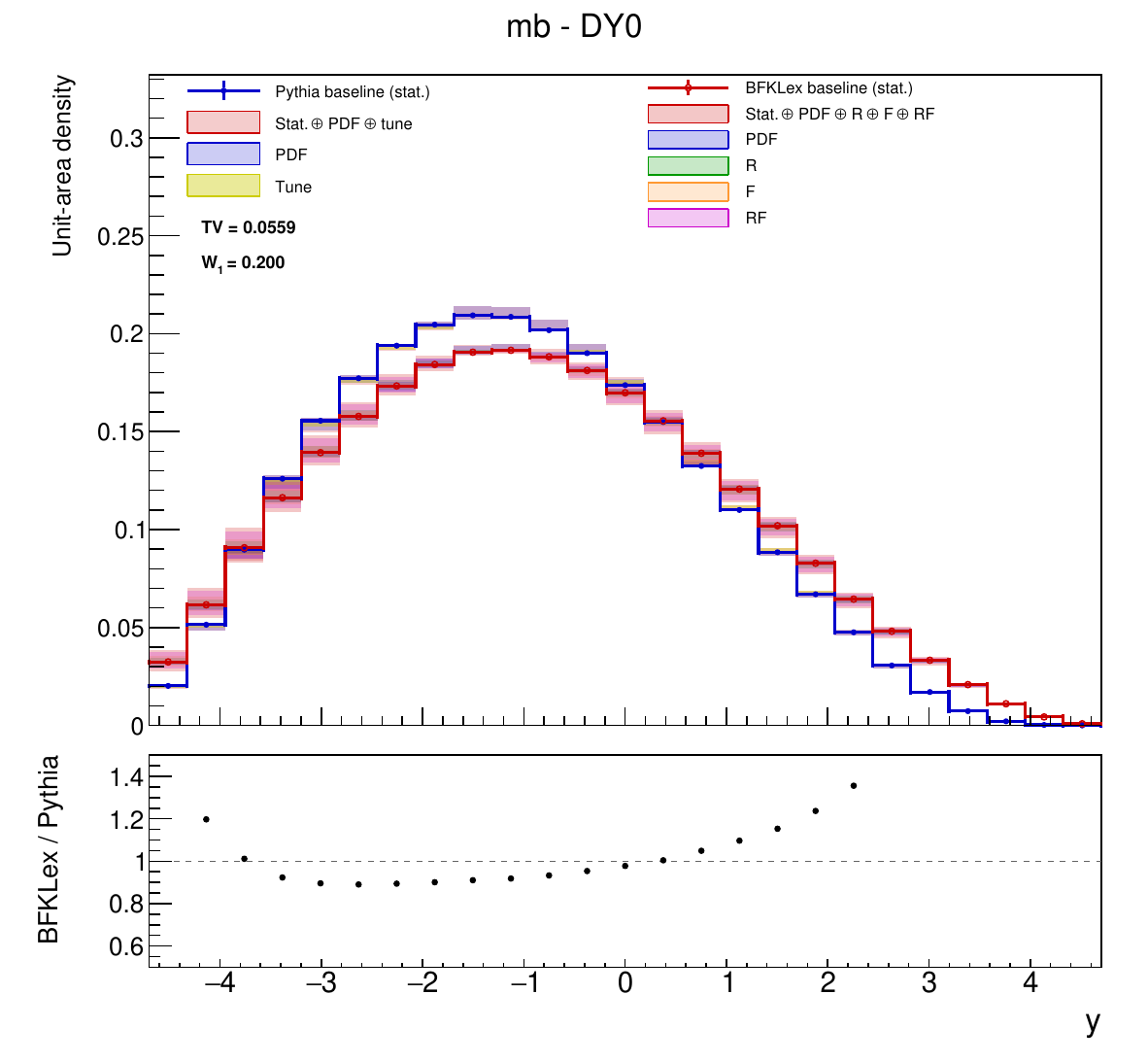}
    \caption{MB (most backward).}
  \end{subfigure}
\caption{\textbf{Across multiplicities and
  $\Delta Y_{\mathrm{MF,MB}}\geq0$.} Normalized MB and MF rapidity spectra
  comparing \textsc{PYTHIA~8.306} and \textsc{BFKLex}, with uncertainty bands as indicated in the legends and baseline ratios in the lower panels.}
  \label{fig:MBMF-Y0}
\end{figure}

\begin{figure}[!htbp]
\ContinuedFloat
\centering
\begin{subfigure}[t]{0.70\textwidth}
    \centering
    \includegraphics[width=\linewidth,trim=0 0 0 23,clip]{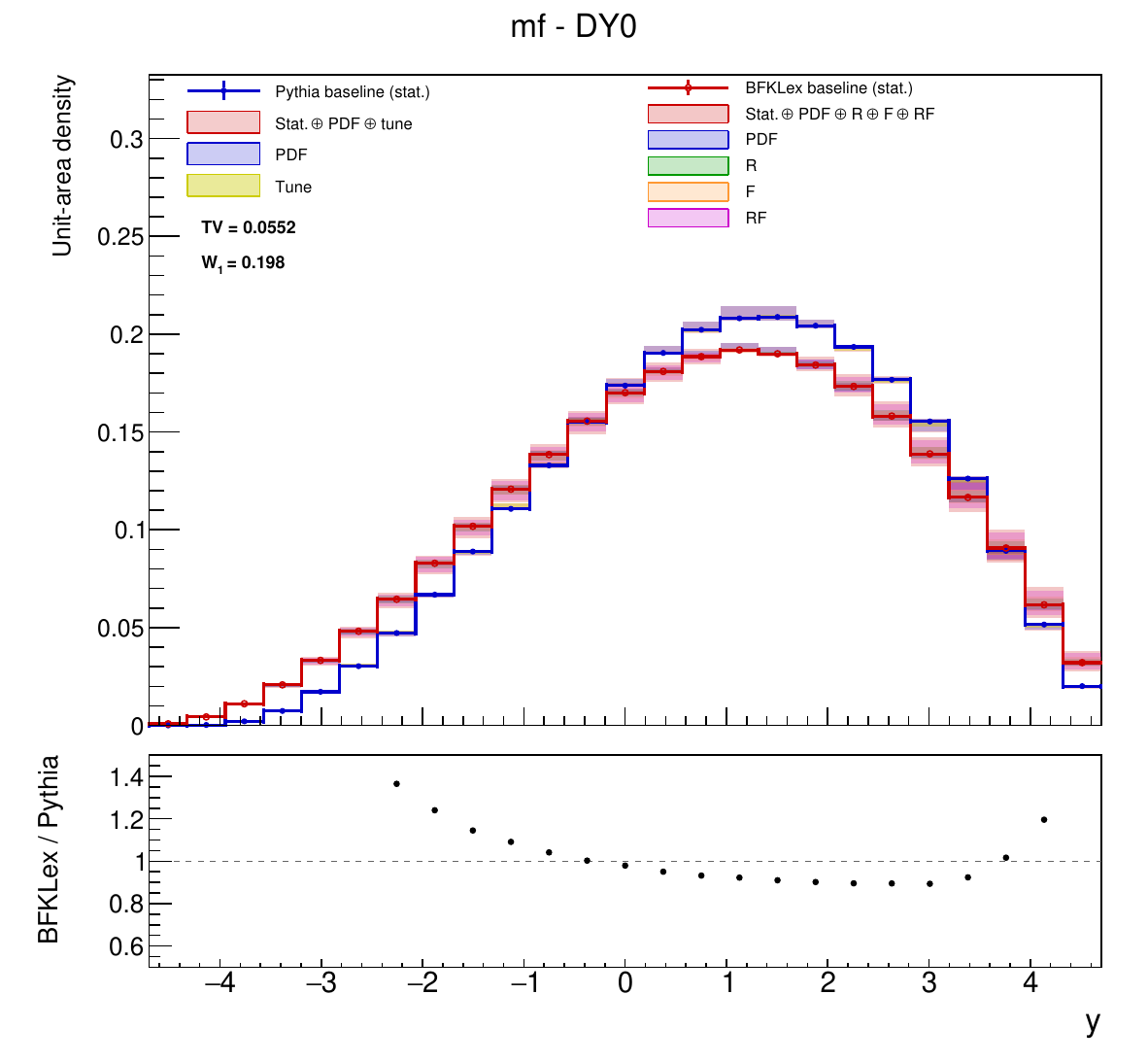}
    \caption{MF (most forward).}
  \end{subfigure}
\captionsetup{labelformat=continuedfigure}
\caption[]{Corresponding forward-rank distribution for the same selection.}
\end{figure}

\begin{samepage}
The SMB and SMF spectra in Fig.~\ref{fig:SMFSMB-Y0} exhibit a larger
separation, with $\mathrm{TV}\simeq0.098$ and $W_1\simeq0.27$:
\textsc{PYTHIA~8.306} favours more concentrated distributions, while
\textsc{BFKLex} gives greater weight to their tails. Thus, restricting the
comparison to the two extremal jets would miss part of the shape
separation revealed by the second-most-backward and second-most-forward
ranks.
\par
\end{samepage}

\begin{figure}[!htbp]
\centering
\begin{subfigure}[t]{0.70\textwidth}
    \centering
    \includegraphics[width=\linewidth,trim=0 0 0 23,clip]{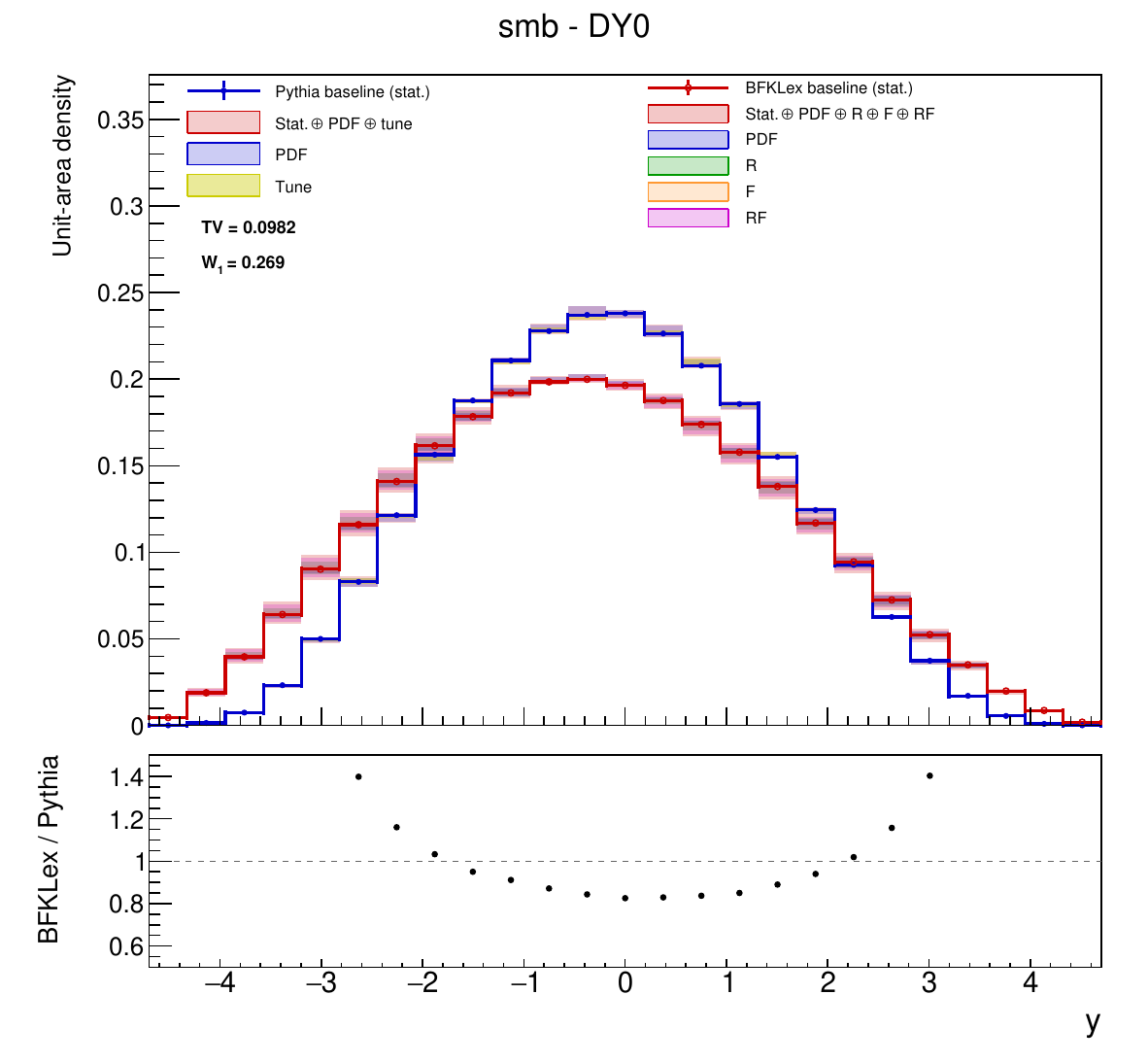}
    \caption{SMB (second-most backward).}
  \end{subfigure}
\caption{\textbf{Across multiplicities and
  $\Delta Y_{\mathrm{MF,MB}}\geq0$.} Normalized SMB and SMF rapidity spectra
  comparing \textsc{PYTHIA~8.306} and \textsc{BFKLex}, with uncertainty bands as indicated in the legends and baseline ratios in the lower panels.}
  \label{fig:SMFSMB-Y0}
\end{figure}

\begin{figure}[!htbp]
\ContinuedFloat
\centering
\begin{subfigure}[t]{0.70\textwidth}
    \centering
    \includegraphics[width=\linewidth,trim=0 0 0 23,clip]{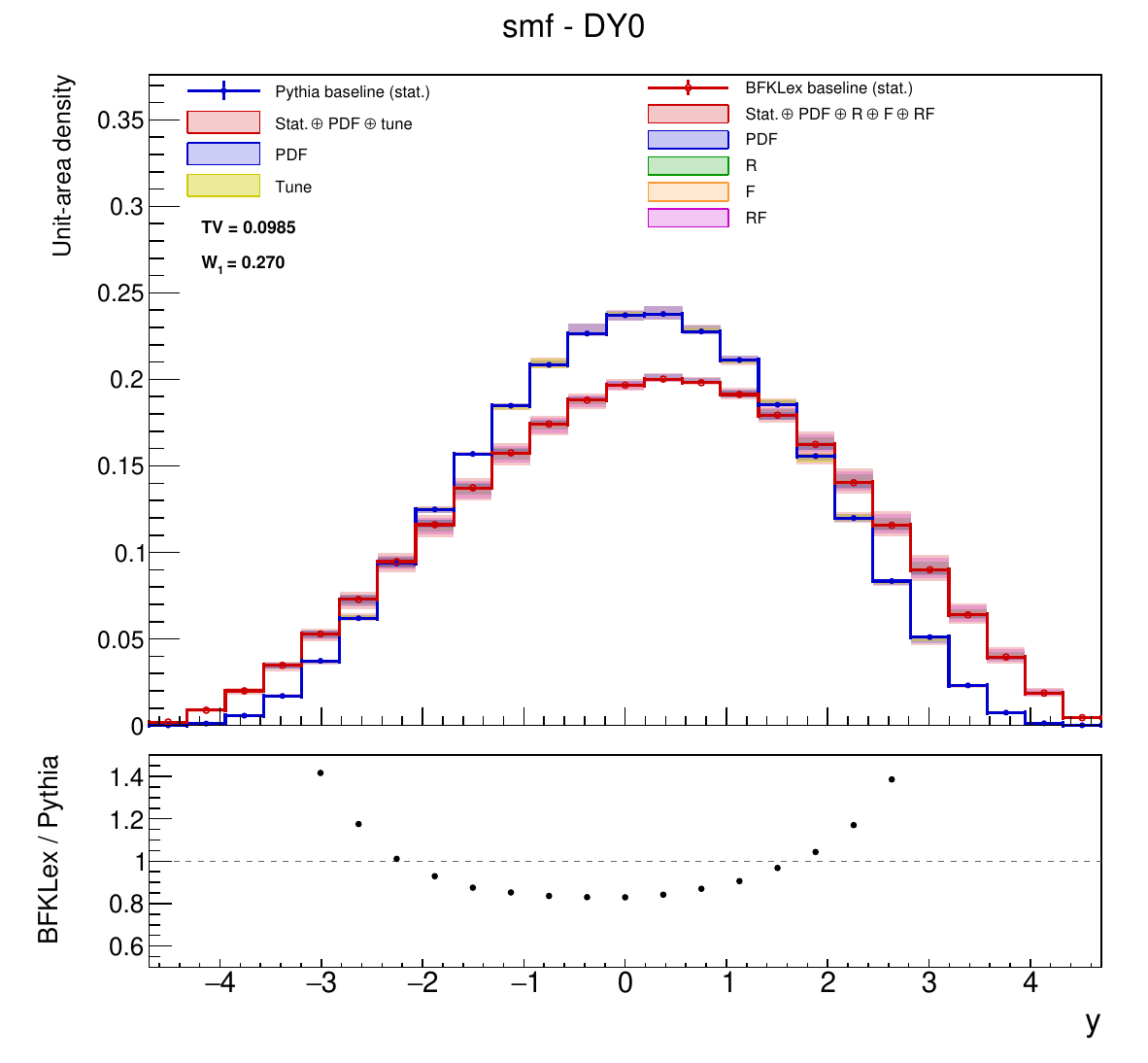}
    \caption{SMF (second-most forward).}
  \end{subfigure}
\captionsetup{labelformat=continuedfigure}
\caption[]{Corresponding forward-rank distribution for the same selection.}
\end{figure}

In several well-populated regions, the differences between the baseline
predictions exceed the widths of the displayed uncertainty bands. This
supports the persistence of the shape separation under the PDF, tune, and
scale variations considered here. The conclusion is local to the regions
and variations shown: the distances summarize the baseline shapes, while
the bands provide a separate assessment of their sensitivity to the varied
inputs.

\subsection{\texorpdfstring{Rapidity-gap requirement
$\Delta Y_{\mathrm{MF,MB}}\geq3$}{Rapidity-gap requirement Delta Y >= 3}}

Figures~\ref{fig:N3-Y3}--\ref{fig:SMFSMB-Y3} repeat the same study after
requiring a boundary-jet separation of at least three units of rapidity.
The same observables and plotting conventions are used to compare the two
generators under this additional rapidity-gap requirement.

The rapidity-gap requirement does not enhance the separation uniformly.
For fixed multiplicity $N=3$, both distances decrease for all three ranks
relative to the selection without an additional gap requirement. In
particular, the outer-rank total-variation distances fall from about $0.12$
to $0.06$, while their Wasserstein distances decrease from about $0.46$
to $0.15$. 

\begin{figure}[!htbp]
\centering
\begin{subfigure}[t]{0.70\textwidth}
    \centering
    \includegraphics[width=\linewidth,trim=0 0 0 23,clip]{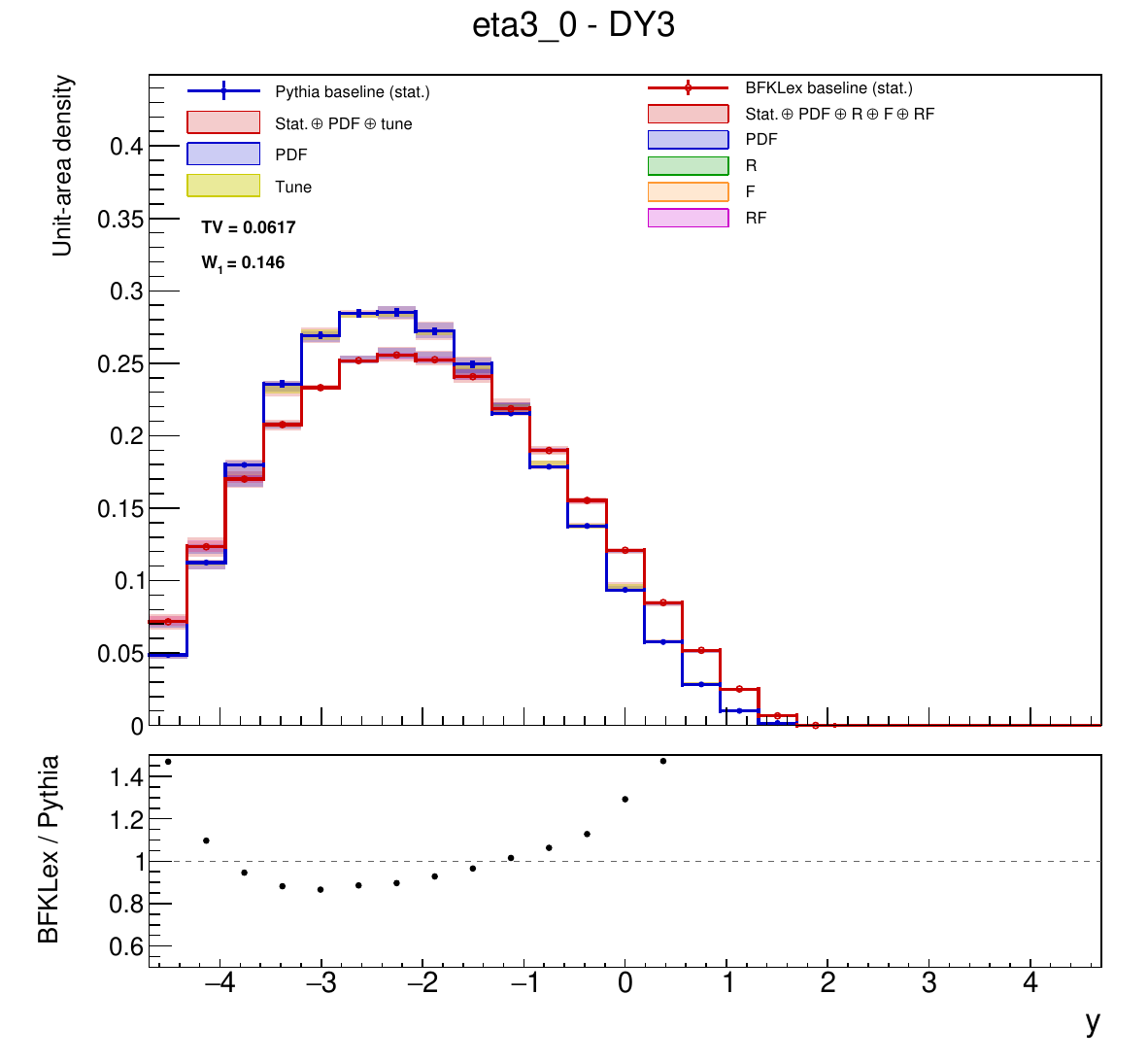}
    \caption{Jet1 (most backward).}
  \end{subfigure}
\caption{\textbf{Fixed multiplicity $N=3$ and
  $\Delta Y_{\mathrm{MF,MB}}\geq3$.} Normalized rapidity spectra for the three ordered jets, comparing \textsc{PYTHIA~8.306} and \textsc{BFKLex}. Bands denote the uncertainty components specified in the legends; lower panels show the \textsc{BFKLex}/\textsc{PYTHIA~8.306} baseline ratio.}
  \label{fig:N3-Y3}
\end{figure}

\begin{figure}[!htbp]
\ContinuedFloat
\centering
\begin{subfigure}[t]{0.70\textwidth}
    \centering
    \includegraphics[width=\linewidth,trim=0 0 0 23,clip]{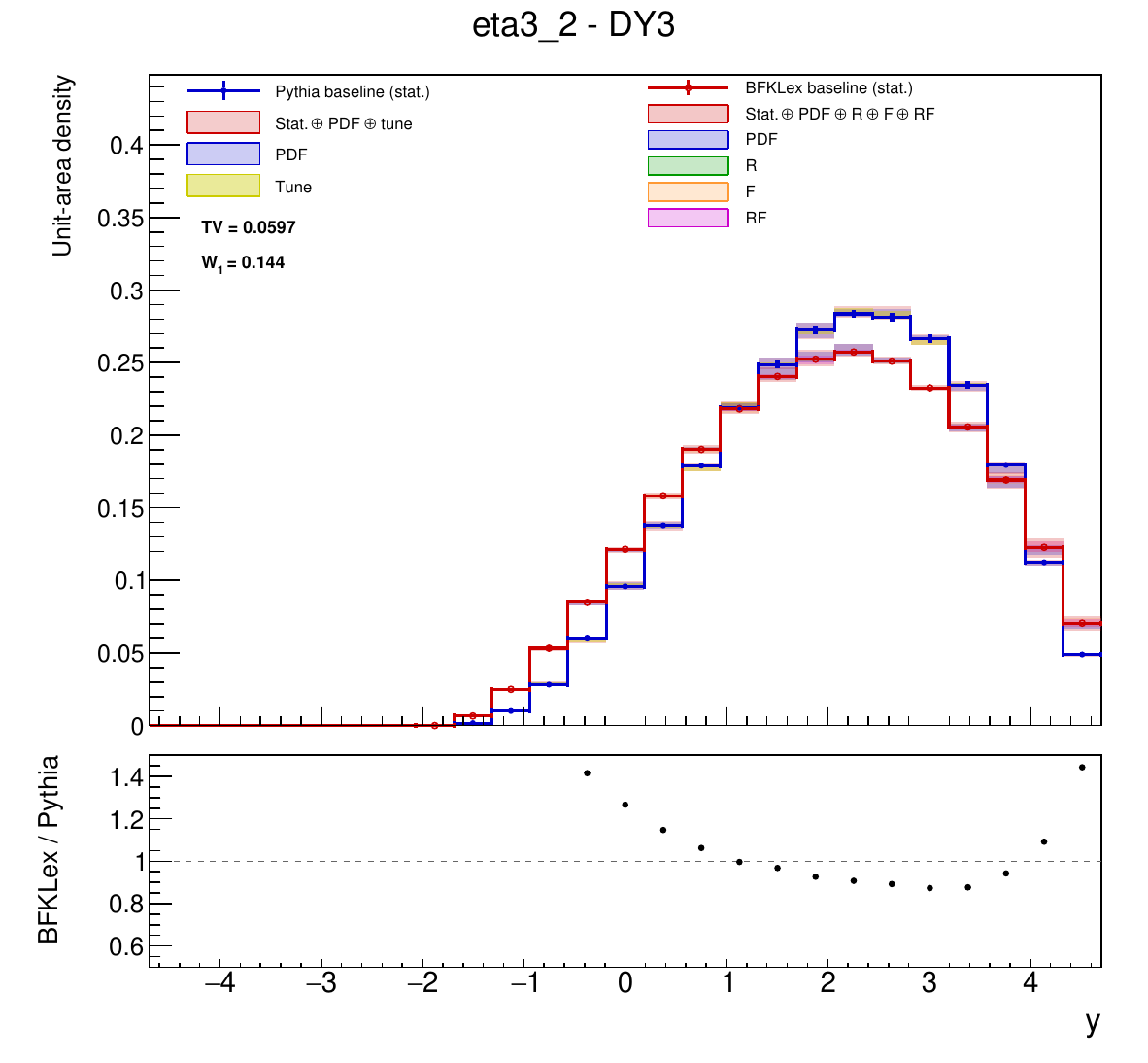}
    \caption{Jet3 (most forward).}
  \end{subfigure}
\captionsetup{labelformat=continuedfigure}
\caption[]{Corresponding forward-rank distribution for the same selection.}
\end{figure}

The central-rank distances also decrease, from $0.0916$ to
$0.0624$ for $\mathrm{TV}$ and from $0.237$ to $0.161$ for $W_1$.
The broader central \textsc{BFKLex} shape remains visible, but the two
predictions are closer than for the unrestricted-gap sample.

\begin{figure}[!htbp]
  \ContinuedFloat
  \centering
\begin{subfigure}[t]{0.70\textwidth}
    \centering
    \includegraphics[width=\linewidth,trim=0 0 0 23,clip]{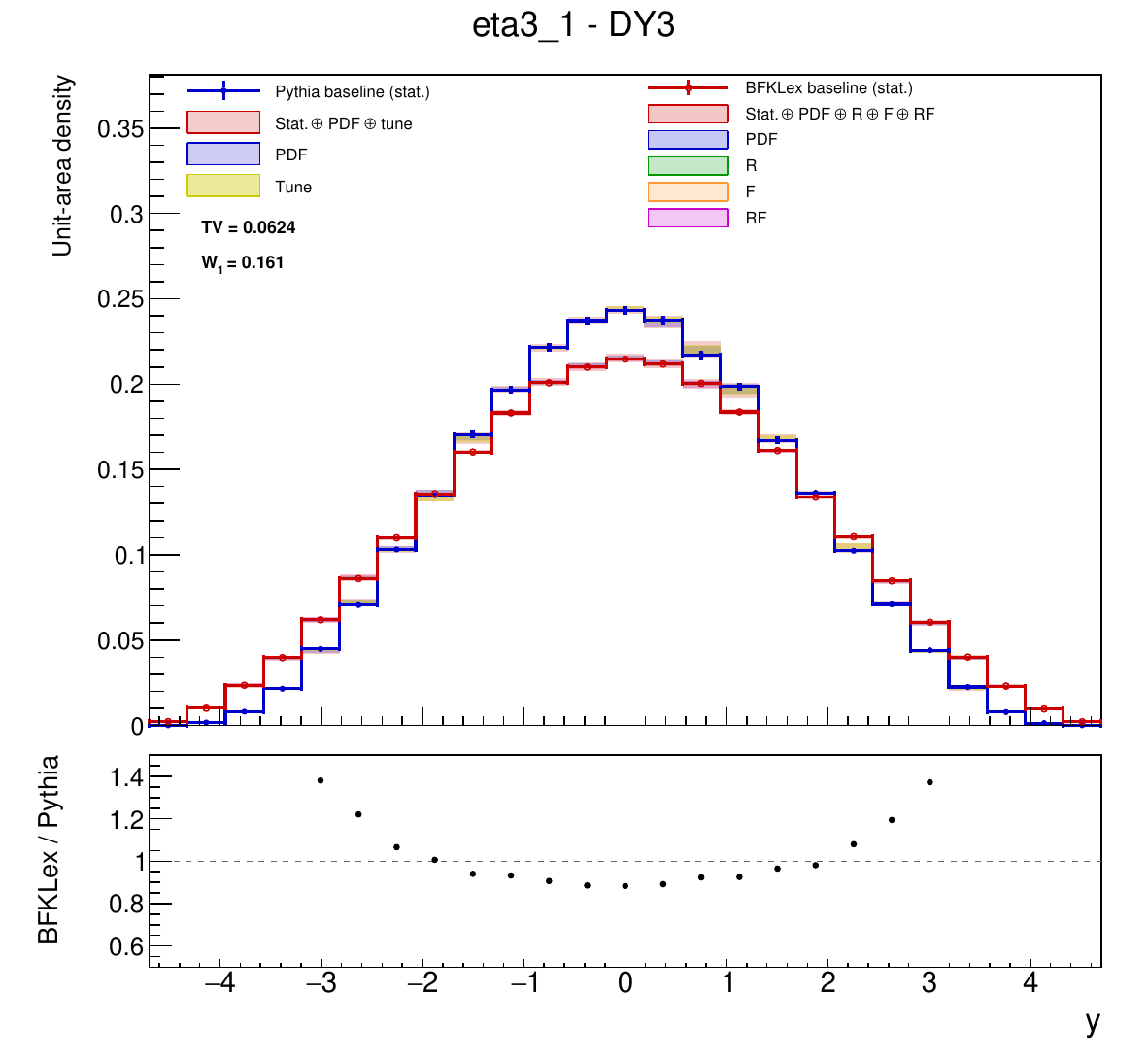}
    \caption{Jet2 (central rank).}
  \end{subfigure}
  \captionsetup{labelformat=continuedfigure}
  \caption[]{Central-rank rapidity spectrum for the same fixed-multiplicity selection.}
\end{figure}

The extremal MB and MF distributions likewise become more similar
(Fig.~\ref{fig:MBMF-Y3}), reaching $\mathrm{TV}\simeq0.029$ and
$W_1\simeq0.05$. 

\begin{figure}[!htbp]
\centering
\begin{subfigure}[t]{0.70\textwidth}
    \centering
    \includegraphics[width=\linewidth,trim=0 0 0 23,clip]{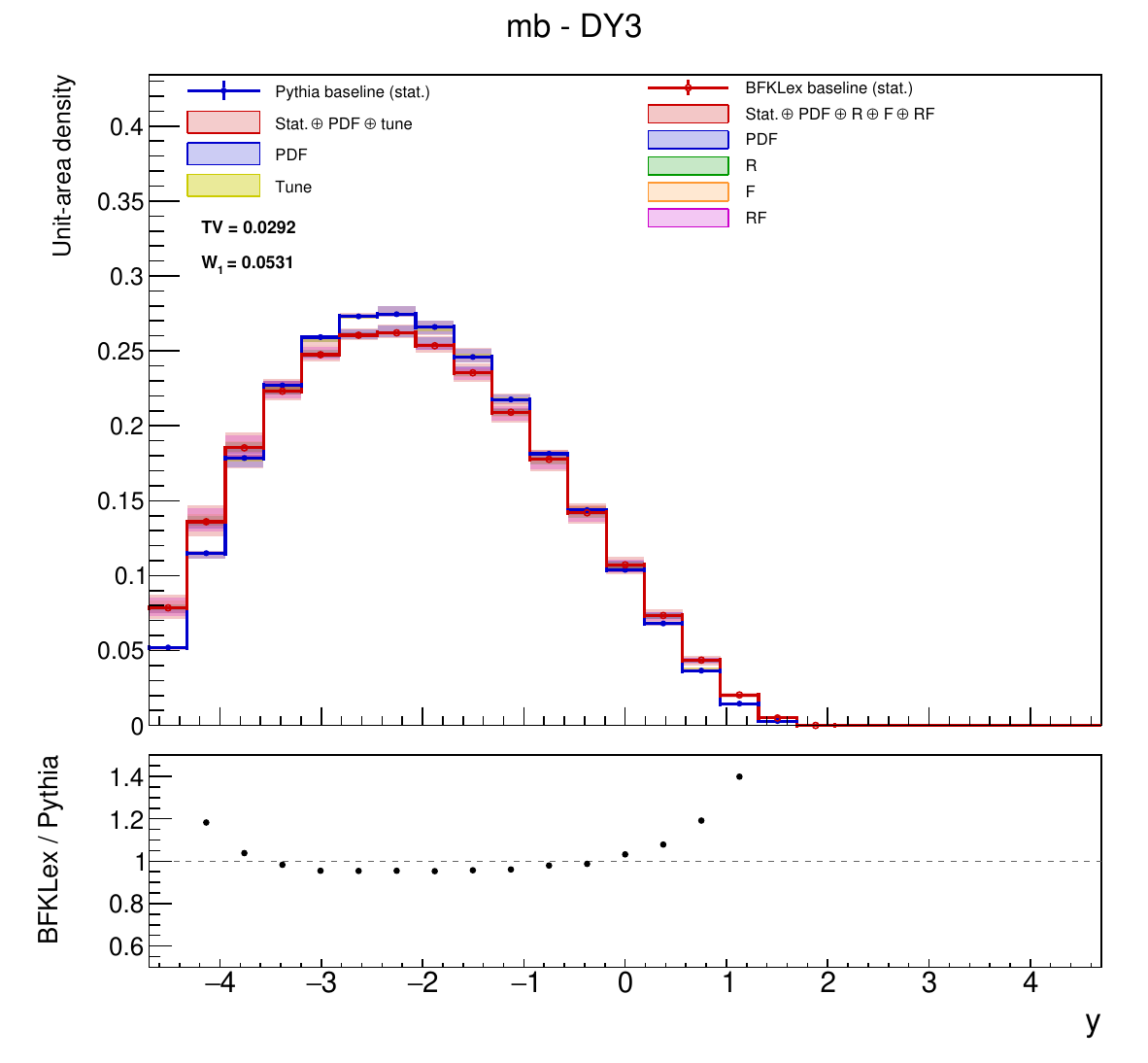}
    \caption{MB (most backward).}
  \end{subfigure}
\caption{\textbf{Across multiplicities and
  $\Delta Y_{\mathrm{MF,MB}}\geq3$.} Normalized MB and MF rapidity spectra
  comparing \textsc{PYTHIA~8.306} and \textsc{BFKLex}, with uncertainty bands as indicated in the legends and baseline ratios in the lower panels.}
  \label{fig:MBMF-Y3}
\end{figure}

\begin{figure}[!htbp]
\ContinuedFloat
\centering
\begin{subfigure}[t]{0.70\textwidth}
    \centering
    \includegraphics[width=\linewidth,trim=0 0 0 23,clip]{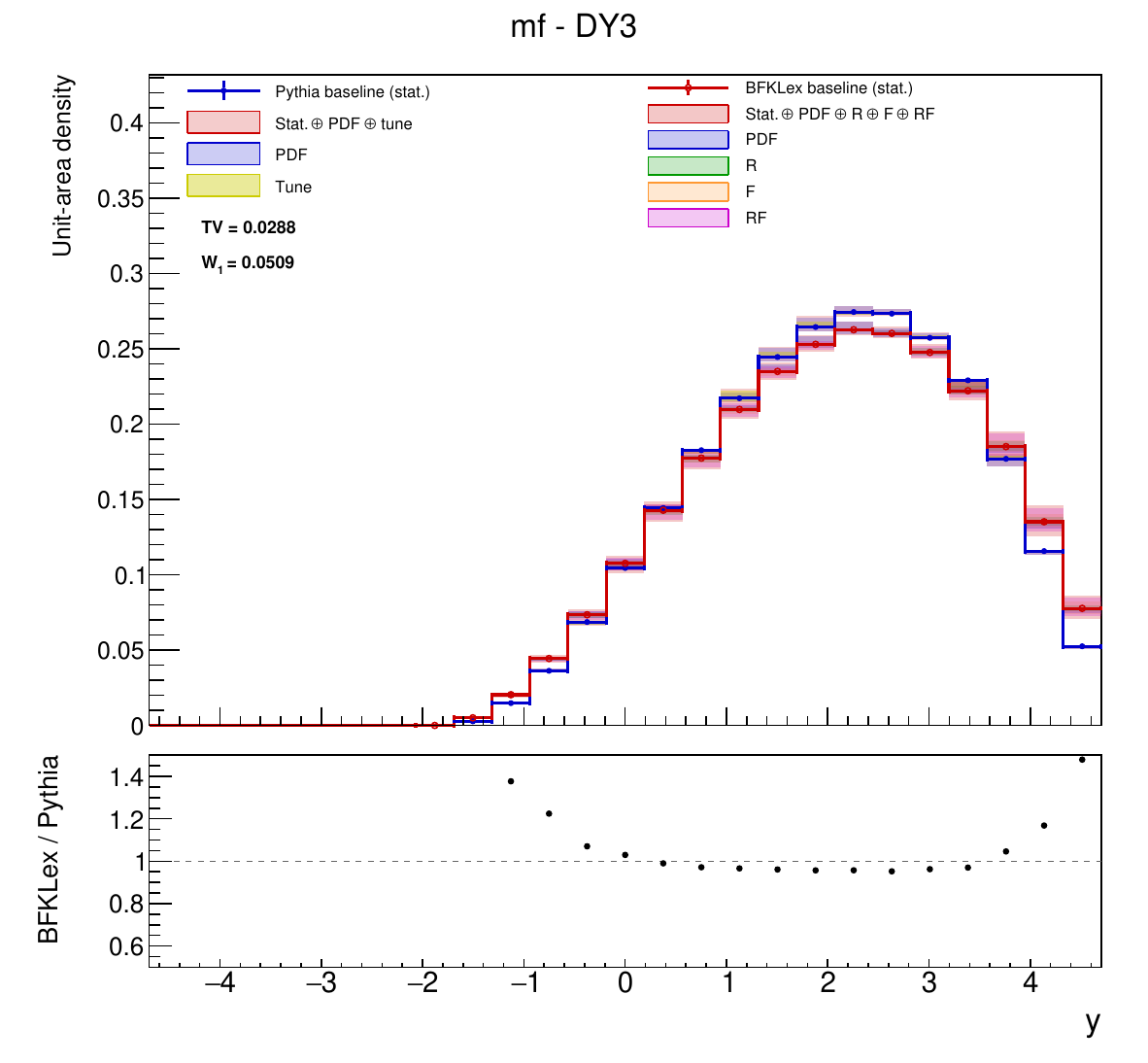}
    \caption{MF (most forward).}
  \end{subfigure}
\captionsetup{labelformat=continuedfigure}
\caption[]{Corresponding forward-rank distribution for the same selection.}
\end{figure}

\begin{samepage}
By contrast, the SMB and SMF distributions show an
increased separation (Fig.~\ref{fig:SMFSMB-Y3}), with
$\mathrm{TV}=0.128$ and $W_1\simeq0.47$. For SMB, the
\textsc{BFKLex} prediction assigns more weight to the backward side than
\textsc{PYTHIA~8.306}; the SMF distribution displays the corresponding
forward-side enhancement. The change therefore involves a redistribution
along the rapidity axis as well as differences in concentration, as
reflected in the simultaneous increase of the two distances. The separation
remains visible relative to the displayed bands in substantial parts of
the populated distributions.
\par
\end{samepage}

\begin{figure}[!htbp]
\centering
\begin{subfigure}[t]{0.70\textwidth}
    \centering
    \includegraphics[width=\linewidth,trim=0 0 0 23,clip]{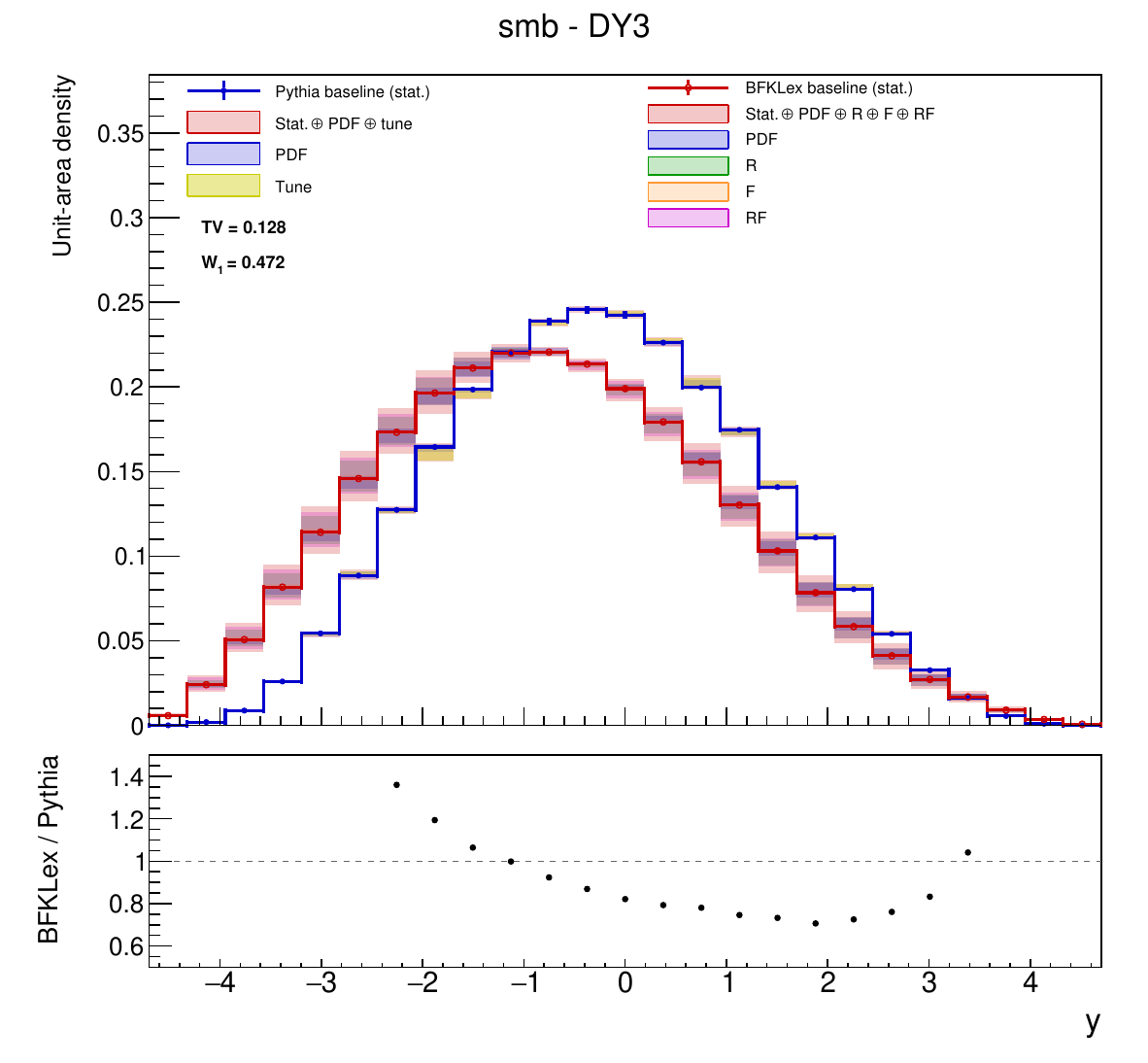}
    \caption{SMB (second-most backward).}
  \end{subfigure}
\caption{\textbf{Across multiplicities and
  $\Delta Y_{\mathrm{MF,MB}}\geq3$.} Normalized SMB and SMF rapidity spectra
  comparing \textsc{PYTHIA~8.306} and \textsc{BFKLex}, with uncertainty bands as indicated in the legends and baseline ratios in the lower panels.}
  \label{fig:SMFSMB-Y3}
\end{figure}

\begin{figure}[!htbp]
\ContinuedFloat
\centering
\begin{subfigure}[t]{0.70\textwidth}
    \centering
    \includegraphics[width=\linewidth,trim=0 0 0 23,clip]{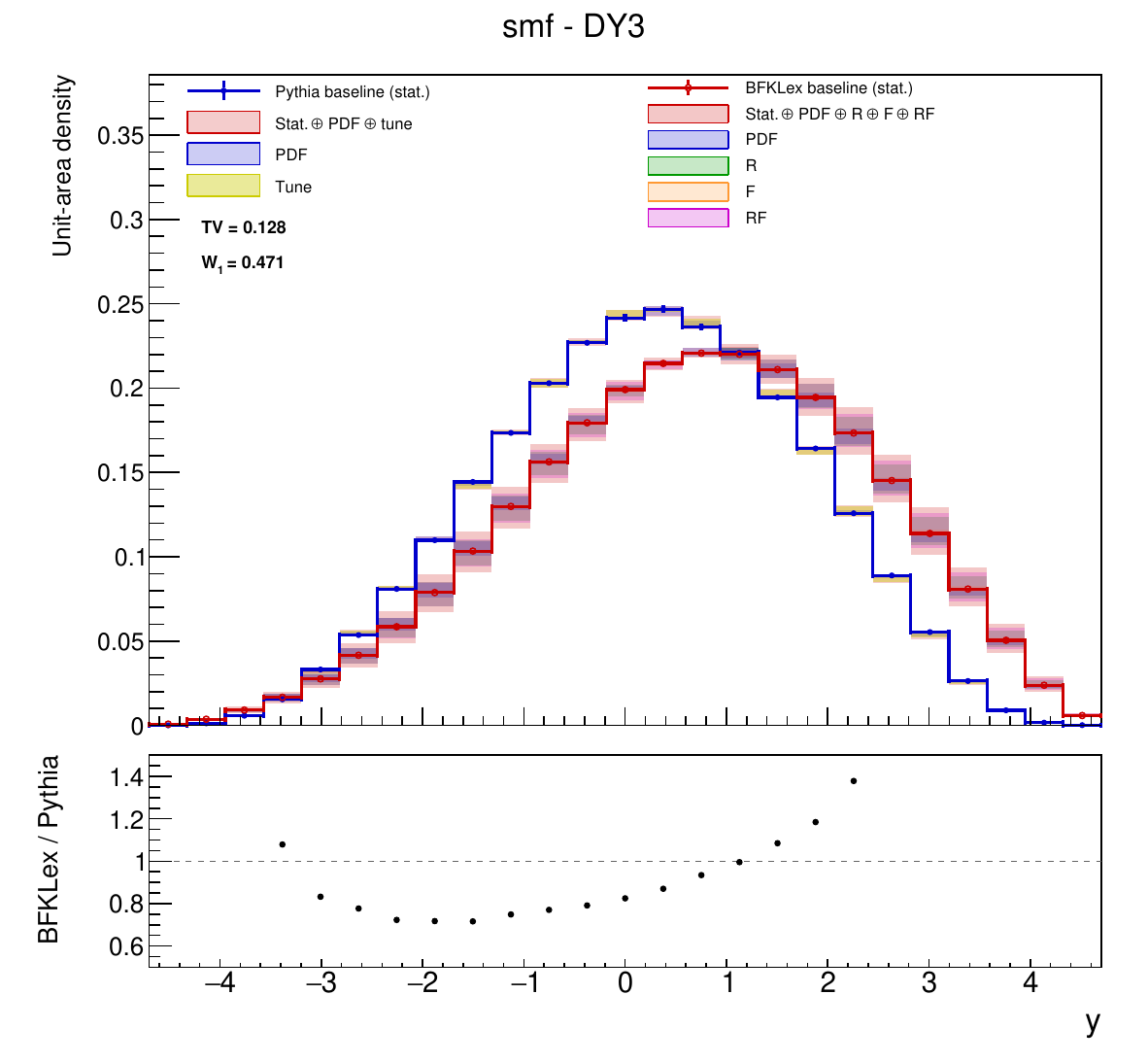}
    \caption{SMF (second-most forward).}
  \end{subfigure}
\captionsetup{labelformat=continuedfigure}
\caption[]{Corresponding forward-rank distribution for the same selection.}
\end{figure}

These results show that the choice of rapidity rank and the gap selection
must be considered together. A possible interpretation is that the gap
requirement constrains the positions of the extremal jets more strongly,
while the second-most-extreme ranks remain sensitive to differences in how
the event is populated. The observed effect is therefore a selective
increase in the discrimination provided by SMB and SMF, rather than a
general increase in separation for every rank.

Table~\ref{tab:shape-distances} summarizes the baseline distances for both
selections, using the values displayed in the comparison plots.

\begin{table}[htbp]
  \centering
  \caption{Shape distances between the normalized \textsc{PYTHIA~8.306} and
  \textsc{BFKLex} baseline predictions for the two rapidity-separation
  selections. $\Delta Y_{\mathrm{MF,MB}}\geq0$ corresponds to no additional
  rapidity-separation requirement.}
  \label{tab:shape-distances}
  \begin{tabular}{lrrrr}
    \toprule
    & \multicolumn{2}{c}{$\Delta Y_{\mathrm{MF,MB}}\geq0$} & \multicolumn{2}{c}{$\Delta Y_{\mathrm{MF,MB}}\geq3$} \\
    \cmidrule(lr){2-3}\cmidrule(lr){4-5}
    Observable & $\mathrm{TV}$ & $W_1$ & $\mathrm{TV}$ & $W_1$ \\
    \midrule
    Jet1 ($N=3$) & 0.121  & 0.467 & 0.0617 & 0.146 \\
    Jet2 ($N=3$) & 0.0916 & 0.237 & 0.0624 & 0.161 \\
    Jet3 ($N=3$) & 0.120  & 0.462 & 0.0597 & 0.144 \\
    MB           & 0.0559 & 0.200 & 0.0292 & 0.0531 \\
    MF           & 0.0552 & 0.198 & 0.0288 & 0.0509 \\
    SMB          & 0.0982 & 0.269 & 0.128  & 0.472 \\
    SMF          & 0.0985 & 0.270 & 0.128  & 0.471 \\
    \bottomrule
  \end{tabular}
\end{table}

\subsection{Comparison with the previous study}

The present results support the qualitative hierarchy observed in
Ref.~\cite{Chachamis:2026ostats}: among the observables defined across
multiplicities, MB and MF show comparatively small shape differences,
whereas SMB and SMF provide greater separation between the generator
predictions. The broader central-rank distribution predicted by
\textsc{BFKLex} at fixed $N=3$ also persists.

Beyond this qualitative agreement, the present analysis evaluates the separation against explicit
PDF and tune variations for \textsc{PYTHIA~8.306} and PDF and scale variations
for \textsc{BFKLex}. The persistence of visible differences relative to
these bands in several populated regions extends the earlier comparison
of normalized shapes by assessing the uncertainty sources considered here.

The event selections differ between the two studies. Reference~\cite{Chachamis:2026ostats}
used a baseline threshold $p_{T,j}>20$~GeV and examined a harder selection
with $p_{T,j}>60$~GeV and $H_{T2}>250$~GeV. Here, the threshold
$p_{T,j}>30$~GeV is combined with the asymmetric boundary-jet selection
of Eq.~\eqref{eq:asymmetric-boundary-cuts}, and the effect of an additional
rapidity-gap requirement is investigated. Numerical changes in the
distances between the two studies should therefore not be interpreted as
an improvement at fixed selection. The new result is the contrasting
response to the gap requirement: the separation decreases for the extremal
MB/MF ranks and for all ranks at fixed $N=3$, but increases for SMB/SMF.
This identifies the second-most-extreme ranks as particularly useful
observables for the $\Delta Y_{\mathrm{MF,MB}}\geq3$ selection and shows
how the gap requirement can help select which rank observables provide the
strongest shape discrimination.

\FloatBarrier

\section{Discussion and outlook}

We have extended the phenomenological study of rank-ordered jet rapidity
observables by comparing normalized \textsc{PYTHIA~8.306} and \textsc{BFKLex}
predictions with an explicit assessment of selected uncertainty sources.
We considered PDF and tune variations for \textsc{PYTHIA~8.306}, and PDF,
factorization-scale, and renormalization-scale variations for
\textsc{BFKLex}. The analysis covers both fixed-multiplicity $N=3$
configurations and rank observables defined across multiplicities. The
total-variation and one-dimensional Wasserstein distances provide
complementary summaries of the differences between the baseline shapes,
quantifying the amount of probability redistributed and its transport
along the rapidity axis, respectively.

The results show that the separation between the two predictions depends
on both the rapidity rank and the event selection. Without an additional
gap requirement, \textsc{PYTHIA~8.306} generally produces more concentrated
rank distributions, while \textsc{BFKLex} assigns greater weight away from
their maxima. Among the observables defined across multiplicities, SMB
and SMF exhibit larger shape differences than MB and MF, consistent with
the qualitative hierarchy found in Ref.~\cite{Chachamis:2026ostats}.
Requiring $\Delta Y_{\mathrm{MF,MB}}\geq3$ reduces both distances for the
extremal MB/MF observables and for all three ranks at fixed $N=3$, but
increases them for SMB/SMF. The gap requirement therefore enhances the
separation selectively, identifying the second-most-extreme ranks as
particularly useful observables for this selection. These findings support
the use of rank distributions to investigate differences between the
\textsc{BFKLex} and \textsc{PYTHIA~8.306} descriptions of high-energy radiation
in the configurations studied here.

In several well-populated regions, the differences between the baseline
predictions remain visible relative to the displayed uncertainty bands.
This supports their robustness under the variations considered. Unit-area normalization removes
common multiplicative rate factors, while retaining effects that change
the shapes. Experimentally, the observables require only reconstructed jet
rapidities and event-by-event ordering. Assessing their achievable precision
requires a dedicated study of trigger efficiencies and effective
luminosity, forward acceptance, detector response, and experimental
systematic uncertainties. 

Natural extensions include next-to-leading-logarithmic BFKL evolution
with the corresponding jet vertices, to assess the perturbative stability
of the observed shape differences, and joint distributions of two ranks,
to explore correlations beyond the one-dimensional projections studied
here. Further studies of event selections and an experimental analysis
would help establish how these observables can complement existing
Mueller--Navelet measurements in probing high-energy QCD radiation.
We therefore strongly encourage the experimental collaborations to consider
measurements of rank-ordered jet rapidity distributions. The present study
is intended to provide a phenomenological foundation for such measurements
and to motivate their inclusion in future analyses of multijet production.

\section*{Acknowledgments}

The work of D. Sunar Cerci and S. Cerci was supported by the Yıldız Technical University Scientific Research Projects Coordination Unit (BAP) under project number FBG-2025-7370.

The work of G. Chachamis was supported by the Fundação para a Ciência
 e a Tecnologia (FCT, Portugal) through project EXPL/FIS-PAR/1195/2021
(\url{http://doi.org/10.54499/EXPL/FIS-PAR/1195/2021}), project
CERN/FIS-PAR/0032/2021
(\url{http://doi.org/10.54499/CERN/FIS-PAR/0032/2021}), and under the
``Investigador FCT'' contract 03216/2017. The work of D. Vaccaro was
supported by the FCT through project EXPL/FIS-PAR/1195/2021 and by the
PT-CERN PhD Grant (Call 2021/2).

A. Sabio Vera acknowledges support from the Spanish Agencia Estatal de
Investigación, grants PID2022-142545NB-C22 and IFT Centro de Excelencia
Severo Ochoa CEX2020-001007-S, funded by MCIN/AEI/10.13039/501100011033
and by ERDF A way of making Europe; and from the European Union's
Horizon 2020 research and innovation programme under grant agreement
No. 824093.

\bibliographystyle{ieeetr}
\bibliography{pheno-paper}

\end{document}